\documentclass[aps,floatfix,amsmath,nofootinbib,amssymb,superscriptaddress,letterpaper,twocolumn]{revtex4-2}

\usepackage{overpic}
\usepackage{amssymb}
\usepackage{indentfirst}
\usepackage{feynmf}
\usepackage{slashed}
\usepackage{cases}
\usepackage{color}
\usepackage{multirow}
\usepackage{epstopdf}
\usepackage{graphicx,color,bm}
\usepackage{booktabs}
\usepackage{float}
\usepackage{amsmath}
\allowdisplaybreaks[4]

\usepackage[colorlinks=true,
            citecolor=green,
            anchorcolor=red,
            menucolor=red,
            linkcolor=red,
            filecolor=red,
            runcolor=red,
            urlcolor=blue]{hyperref}

\begin{document}

\title{Gluon TMDs for tensor polarized deuteron in a spectator model}

\author{Xiupeng Xie}\affiliation{School of Physics, Southeast University, Nanjing
211189, China}

\author{Dian-Yong Chen}
\email{chendy@seu.edu.cn}
\affiliation{School of Physics, Southeast University, Nanjing
211189, China}

\author{Zhun Lu}
\email{zhunlu@seu.edu.cn}
\affiliation{School of Physics, Southeast University, Nanjing 211189, China}

\begin{abstract}
We present a model calculation of the transverse-momentum-dependent distributions (TMDs) for gluons in a tensor-polarized deuteron. Our model is based on the assumption that an on-shell deuteron can emit a time-like off-shell gluon, while the remaining system is treated as a single on-shell spectator particle whose mass can take on a continuous range of real values, described by a spectral function. For spin-1 hadrons, the polarization is characterized not only by a spin vector $S$ but also by a symmetric traceless spin tensor $T$. The deuteron-gluon-spectator coupling is described by an effective vertex containing three form factors. We obtain analytical expressions for thirteen T-even gluon TMDs. We also provide numerical results for the $x$-dependence and $\bm{k}_T$-dependence of these TMDs. Our analysis reveals non-negligible results of these gluon TMDs, especially for tensor-polarized hadrons, which could potentially be explored in future experimental measurements.

\end{abstract}

\maketitle

\section{Introduction}

Transverse-momentum-dependent parton distributions (TMDs) are fundamental quantities that encode novel information about the internal momentum and spin structure of hadrons, extending beyond the knowledge obtained from high-precision measurements of conventional parton distribution functions (PDFs). TMDs describe the parton distributions as functions of both the partonic transverse momentum $\bm{k}_T$ and the longitudinal momentum fraction $x$. Their significance stems from their unique sensitivity to the dynamics of color charges in QCD, as the transverse momentum dependence is intimately connected with gauge links and, consequently, with the non-perturbative behavior of color fields, which are otherwise difficult to access due to color confinement.

Due to their intrinsically non-perturbative nature, TMDs cannot be directly computed from first principles using perturbative QCD alone. Describing hadron structure thus requires solving QCD in the non-perturbative regime. The state-of-the-art first-principle tool for this purpose is lattice QCD (LQCD), which has achieved impressive progress in recent decades~\cite{Rothe:1992nt,Gattringer:2010zz,Ji:2013dva,Ji:2014gla}. Phenomenological models~\cite{Brodsky:2002cx,Collins:2002kn,Metz:2002iz,Collins:2004nx,Kundu:2001pk,Goeke:2003az,Meissner:2009ww}, which serve as essential tools for gaining insights into hadron structure, play a complementary role to lattice studies. They are typically employed to provide insights into the underlying dynamics and offering predictions for quantities that are currently challenging to access via lattice simulations. While TMDs have been extensively investigated for spin-1/2 nucleons~\cite{Metz:2016swz,Bacchetta:2006tn,Mulders:2000sh}, studies of TMDs and structure functions for spin-1 hadrons remain relatively underdeveloped.

On the experimental side, no new measurements of polarized structure functions of spin-1 hadrons have been reported since the HERMES measurement of the $b_1(x)$ function for the deuteron in 2005~\cite{HERMES:2005pon}. However, this situation is expected to change with several upcoming or proposed experimental programs, including those at the JLab~\cite{Maxwell:2018gci}, the Fermilab~\cite{Keller:2022abm}, the NICA~\cite{Arbuzov:2020cqg}, the LHC-spin~\cite{Aidala:2019pit}, and the electron-ion colliders (EIC,EicC)~\cite{AbdulKhalek:2021gbh,Anderle:2021wcy}. Theoretically, the $b_1(x)$ structure function can be estimated, for example, using a convolution model of nucleon light-cone momentum distributions. Predictions from such simple models, however, often deviate significantly from the HERMES data, highlighting the need for more sophisticated theoretical approaches.

A particularly intriguing observable is the gluon structure function $\Delta_T g(x)$~\cite{Jaffe:1989xy}, which corresponds to the PDF $h_{1TT}(x)$. This function is associated with the double-helicity flip amplitude in processes involving hadrons with spin $\geq 1$. Specifically, $h_{1TT}(x)$ describes the distribution of linearly polarized gluons within a transversely tensor-polarized target. It is sometimes referred to as ``gluon transversity", a term that can be misleading, as it suggests a transverse polarization of the gluon itself, whereas it actually specifies the polarization state of the target hadron. The experimental observation of a non-zero $\Delta_T g(x)$ would be a clear signal of non-nucleonic degrees of freedom within the deuteron, as such a distribution cannot arise from a simple system of bound nucleons. Therefore, measurements of unique spin-1 observables like $b_1(x)$ and $\Delta_T g(x)$ offer a potential gateway to unraveling new aspects of hadron physics in the nuclear domain.

In this work, we calculate the leading-twist T-even gluon TMDs for a tensor-polarized deuteron within a spectator model framework. In this model, the deuteron is assumed to emit a time-like, off-shell gluon, while the remaining system is effectively treated as a single on-shell spectator particle. The mass of this spectator is not fixed but instead take a continuous range of values, as described by a spectral function. To fully characterize the polarization of a spin-1 hadron, its density matrix involves not only a spin vector $S^\mu$ but also a symmetric, traceless rank-2 spin tensor $T^{\mu\nu}$. The deuteron-gluon-spectator vertex is modeled via an effective coupling that incorporates three independent form factors. This spectator approach has been successfully applied in previous studies to compute quark TMDs for nucleons~\cite{Jakob:1997wg,Brodsky:2002cx,Gamberg:2003ey,Bacchetta:2003rz,Bacchetta:2008af,Bacchetta:2010si} and pions~\cite{Lu:2004hu,Meissner:2008ay,Ma:2019agv}, as well as gluon TMDs and TMD fragmentation functions for the proton~\cite{Lu:2016vqu,Bacchetta:2020vty,Bacchetta:2024fci,Xie:2022lra,Xie:2024cnv}. Here, we extend this formalism to the case of a tensor-polarized deuteron. At leading twist, the T-even gluon TMDs consist of the unpolarized distribution $f_1$, three vector-polarized distributions ($h_1^\perp$, $g_1$, $g_{1T}$), and nine tensor-polarized distributions ($f_{1LL}$, $f_{1LT}$, $f_{1TT}$, $h_{1LL}^{\perp}$, $h_{1LT}$, $h_{1LT}^{\perp}$, $h_{1TT}$, $h_{1TT}^{\perp}$, $h_{1TT}^{\perp\perp}$). The model parameters are determined by fitting to the unpolarized gluon TMD $f_1$, ensuring a realistic description of the non-polarized baseline.

The structure of this paper is as follows. In Sec.~\ref{section2}, we present the spectator model formalism for gluon TMDs. We derive the tree-level gluon-gluon correlator and extract the leading-twist T-even TMDs via appropriate projection operators. In Sec.~\ref{section3}, we determine the model parameters by fitting to the unpolarized TMD $f_1$, and then present numerical predictions for all T-even gluon TMDs. We summarize our findings and conclude in Sec.~\ref{section4}.

\section{Analytic calculation of the T-even TMDs of tensor polarized deuteron}\label{section2}

We work with the light-cone vectors $n_+^\mu \equiv [1,0,\bm{0}_T]$ and $n_-^\mu \equiv [0,1,\bm{0}_T]$, satisfying $n_+ \cdot n_- = 1$ and $n_\pm^2 = 0$. Any four-vector $a^\mu$ admits the light-cone decomposition $a^\mu = a^+ n_+^\mu + a^- n_-^\mu + a_T^\mu$, with $a^\pm \equiv a \cdot n_\mp$ and $a_T^\mu \equiv (0,0,\bm{a}_T)$ denoting its transverse component. In a frame where the hadron carries no transverse momentum, the hadron momentum $P^\mu$ and the parton momentum $k^\mu$ are given by
\begin{align}
P^\mu=&P^+ n_+^\mu + \frac{M^2}{2P^+} n_-^\mu \,,\\
k^\mu=&xP^+ n_+^\mu + \frac{k^2+\bm{k}_T^2}{2xP^+} n_-^\mu +\bm{k}_T \,,
\end{align}
where $M$ is the hadron mass and $x=k^+/P^+$ is the longitudinal momentum fraction carried by the parton.

For a spin-$J$ particle, the polarization density matrix is parameterized in terms of irreducible spin tensors up to rank $2J$. 
Following Ref.~\cite{Bacchetta:2000jk}, a complete description of the polarization of a spin-1 hadron requires both a spin vector $S^\mu$ and a symmetric traceless rank-2 spin tensor $T^{\mu\nu}$. Subject to the constraints $P \cdot S = 0$ and $P_\mu T^{\mu\nu} = 0$, they are decomposed as
\begin{align}
S^\mu=&S_L \frac{P^+}{M} n_+^\mu-S_L\frac{M}{2P^+} n_-^\mu+ S_T^\mu\,,\\
T^{\mu \nu}=&\frac{1}{2}\Bigg[\frac{4}{3}S_{LL}\frac{\left(P\cdot n_-\right)^2}{M^2} n_+^\mu n_+^\nu-\frac{2}{3}S_{LL}\left(n_+^{\left\{\mu\right.}n_-^{\left.\nu \right\}}-g_T^{\mu \nu}\right)\notag\\
&+\frac{1}{3}S_{LL}\frac{M^2}{\left(P \cdot n_-\right)^2}n_-^\mu n_-^\nu+\frac{P\cdot n_-}{M}n_+^{\left\{\mu\right.} S_{LT}^{\left.\nu\right\}}\notag\\
&-\frac{M}{2P\cdot n_-}n_-^{\left\{\mu\right.} S_{LT}^{\left.\nu\right\}}+S_{TT}^{\mu\nu}\Bigg]\,,
\end{align}
where $g_T^{\mu\nu} = g^{\mu\nu} - n_+^{\{\mu} n_-^{\nu\}}$ is the symmetric transverse metric tensor, and the notation $a^{\{\mu} b^{\nu\}} \equiv a^\mu b^\nu + a^\nu b^\mu$ denotes symmetrization of the indices.

The polarization sum for a spin-1 hadron can be expressed in terms of $S^\mu$ and $T^{\mu\nu}$ as
\begin{align}
&\epsilon^{*\mu}\left(P,\lambda\right) \epsilon^\nu \left(P,\lambda\right)\notag\\
=&-\frac{1}{3}\left(g^{\mu \nu}-\frac{P^\mu P^\nu}{M^2}\right)+\frac{i}{2M}\epsilon^{\mu \nu \alpha \beta}P_\alpha S_\beta-T^{\mu \nu}\,.
\end{align}

The gauge-invariant gluon-gluon correlator for spin-1 hadrons is expressed as~\cite{Mulders:2000sh,Meissner:2007rx}
\begin{align}
\Phi&^{\mu \nu;\rho \sigma}\left(x,\bm{k}_T;S,T\right)=\frac{1}{xP^+}\int \frac{d\xi^- d \bm{\xi}_T}{(2\pi)^3} e^{ik\cdot \xi} \notag\\
&\times \left. \left\langle P,S,T\left|F^{\mu \nu}\left(0\right) U_{[0,\xi]} F^{\rho \sigma}\left(\xi\right) U_{[\xi,0]}^\prime \right|P,S,T\right\rangle\right|_{\xi^+=0}\,,\label{eq:phi0}
\end{align}
where a summation over color indices is implicit, and the two process-dependent gauge links $U_{[0,\xi]}$ and $U_{[\xi,0]}^\prime$ ensure color gauge invariance.

At leading twist, we focus on the correlator $\Phi^{+i,+j} \equiv \Phi^{ij}$ with $i,j$ being transverse spatial indices. This correlator can be parametrized in terms of the leading-twist gluon TMDs as~\cite{Boer:2016xqr}
\begin{align}
\Phi^{ij}_U \left(x,\bm{k}_T\right)=&\frac{1}{2}\left[ -g_T^{ij} f_1\left(x,\bm{k}_T^2\right)+\frac{k_T^{ij}}{M^2}h_1^\perp \left(x,\bm{k}_T^2\right)\right]\,,\label{eq:phiU}\\
\Phi^{ij}_L \left(x,\bm{k}_T\right)=&\frac{1}{2}\Bigg[ i \epsilon_T^{ij}S_L g_1\left(x,\bm{k}_T^2\right)\notag\\
&+\frac{\epsilon_{T \alpha}^{\left\{i\right.} k_T^{\left.j\right\}\alpha } S_L}{2M^2} h_{1L}^\perp \left(x,\bm{k}_T^2\right)\Bigg]\,,\\
\Phi^{ij}_T \left(x,\bm{k}_T\right)=&\frac{1}{2}\Bigg[-\frac{g_T^{ij}\epsilon_T^{S_T k_T}}{M} f_{1T}^\perp \left(x,\bm{k}_T^2\right)\notag\\
&+\frac{i \epsilon_T^{ij} \bm{k}_T \cdot \bm{S}_T}{M}g_{1T}\left(x,\bm{k}_T^2\right) \notag\\ 
&-\frac{\epsilon_T^{k_T \left\{i\right.} S_T^{\left.j\right\}}+\epsilon_T^{S_T \left\{i\right.} k_T^{\left.j\right\}}}{4M} h_1 \left(x,\bm{k}_T^2\right)\notag\\
&-\frac{\epsilon_{T \alpha}^{\left\{i\right.} k_T^{\left.j\right\}\alpha S_T}}{2M^3} h_{1T}^\perp \left(x,\bm{k}_T^2\right)\Bigg]\,,\\
\Phi^{ij}_{LL} \left(x,\bm{k}_T\right)=&\frac{1}{2}\Bigg[-g_T^{ij} S_{LL} f_{1LL}\left(x,\bm{k}_T^2\right)\notag\\
&+\frac{k_T^{ij} S_{LL}}{M^2}h_{1LL}^\perp \left(x,\bm{k}_T^2\right)\Bigg]\,,\\
\Phi^{ij}_{LT} \left(x,\bm{k}_T\right)=&\frac{1}{2}\Bigg[-\frac{g_T^{ij} \bm{k}_T \cdot \bm{S}_{LT}}{M} f_{1LT}\left(x,\bm{k}_T^2\right)\notag\\
&+\frac{i \epsilon_T^{ij} \epsilon_T^{S_{LT}k_T}}{M}g_{1LT}\left(x,\bm{k}_T^2\right)\notag\\
&+\frac{S_{LT}^{\left\{i\right.} k_T^{\left.j\right\}}}{M} h_{1LT}\left(x,\bm{k}_T^2\right)\notag\\
&+\frac{k_T^{ij\alpha} S_{LT\alpha}}{M^3} h_{1LT}^\perp \left(x,\bm{k}_T^2\right)\Bigg]\,,\\
\Phi^{ij}_{TT} \left(x,\bm{k}_T\right)=&\frac{1}{2}\Bigg[-\frac{g_T^{ij}k_T^{\alpha \beta} S_{TT\alpha \beta}}{M^2} f_{1TT}\left(x,\bm{k}_T^2\right)\notag\\
&+\frac{i \epsilon_T^{ij} \epsilon_{T\gamma}^{\beta} k_T^{\gamma \alpha} S_{TT\alpha \beta}}{M^2} g_{1TT}\left(x,\bm{k}_T^2\right)\notag\\
&+S_{TT}^{ij} h_{1TT}\left(x,\bm{k}_T^2\right)\notag\\
&+\frac{S_{TT \alpha}^{\left\{i\right.} k_T^{\left.j\right\}\alpha}}{M^2}h_{1TT}^\perp \left(x,\bm{k}_T^2\right)\notag\\
&+\frac{k_T^{ij \alpha \beta} S_{TT\alpha \beta}}{M^4} h_{1TT}^{\perp \perp}\left(x,\bm{k}_T^2\right)\Bigg]\,.\label{eq:phiTT}
\end{align}
Here, $\epsilon_T^{ij}=\epsilon^{n_+ n_- ij}=\epsilon^{-+ij}$ is the anti-symmetric transverse tensor, with the nonzero components $\epsilon_T^{12}=1$. The symmetric traceless tensors $k_T^{i_1 ... i_n}$ are defined as:
\begin{align}
k_T^{ij}\equiv& k_T^i k_T^j +\frac{1}{2} \bm{k}_T^2 g_T^{ij}\,,\\
k_T^{ijk}\equiv& k_T^i k_T^j k_T^k +\frac{1}{4}\bm{k}_T^2 \left( g_T^{ij}k_T^k+g_T^{ik}k_T^j+g_T^{jk}k_T^i\right)\,,\\
k_T^{ijkl}\equiv& k_T^i k_T^j k_T^k k_T^l +\frac{1}{6} \bm{k}_T^2 \left(g_T^{ij}k_T^{kl}+g_T^{ik}k_T^{jl}+g_T^{il}k_T^{jk}\right.\notag\\
&\left.+g_T^{jk}k_T^{il}+g_T^{jl}k_T^{ik}+g_T^{kl}k_T^{ij}\right)\notag\\
&-\frac{1}{8} \bm{k}_T^4\left(g_T^{ij} g_T^{kl}+g_T^{ik} g_T^{jl}+g_T^{il} g_T^{jk}\right)\,,
\end{align}
satisfying the traceless conditions:
\begin{align}
    g_{Tij}k_T^{ij}=g_{Tij}k_T^{ijk}=g_{Tij}k_T^{ijkl}=0\,.
\end{align}

\begin{table}[H]
\centering
\caption{The columns indicate the gluon polarization, unpolarized (U), circularly polarized(Circ), linearly polarized (Lin). The rows indicate the hadron polarization, unpolarized (U), vector polarized (L or T), and tensor polarized (LL, LT, or TT).}\label{table:func}
    \setlength{\tabcolsep}{0.4cm}{
    \begin{tabular}{cccc}
    \hline
    $H \backslash g$ & U & Circ & Lin \\
    \hline
    U & $f_1$ &  & $h_1^\perp$ \\

    L &  & $g_1$ & $h_{1 L}^{\perp}$ \\

    T & $f_{1 T}^{\perp}$ & $g_{1 T}$ & $h_{1} \quad h_{1 T}^{\perp}$ \\

    LL&$f_{1LL}$&   &$h_{1LL}^\perp$\\

    LT&$f_{1LT}$&$g_{1LT}$&$h_{1LT} \quad h_{1LT}^\perp$\\

    TT&$f_{1TT}$&$g_{1TT}$&$h_{1TT} \quad h_{1TT}^{\perp} \quad h_{1TT}^{\perp \perp}$\\
    \hline
    \end{tabular}}
\end{table}

Table~\ref{table:func} summarizes the gluon TMDs appearing in Eqs.~(\ref{eq:phiU})-(\ref{eq:phiTT}), classified according to hadron and parton polarizations. Among the nineteen functions, six are odd under time reversal (T-odd): $f_{1T}^\perp$, $h_{1L}^\perp$, $h_1$, $h_{1T}^\perp$, $g_{1LT}$, and $g_{1TT}$. In this work, we restrict ourselves to tree-level contributions and neglect effects arising from gauge links and their process dependence. At tree level, the correlator does not yield nonvanishing T-odd TMDs, as these typically require the interference of two amplitudes with distinct imaginary parts. To generate the necessary imaginary parts in the scattering amplitude, one must take into account gluon loop diagrams, including those arising from the gauge-link expansion.

In the spectator model, the deuteron state  $\left|\left.P,S,T\right\rangle\right.$, characterized by momentum $P$, spin vector $S$, and spin tensor $T$, is assumed to split into an emitted gluon with momentum $k$, while the remaining system is treated as a single on-shell spectator particle of spin-1, denoted by $\left|\left.P-k\right\rangle\right.$, with momentum $P-k$ and mass $M_S$.

The tree-level gluon-gluon correlator in this model is given by
\begin{align}
\Phi^{ij}&\left(x,\bm{k}_T;S,T\right)=\frac{1}{xP^+}\frac{1}{\left(2\pi\right)^3} \frac{1}{2\left(1-x\right)P^+}\notag\\
&\times \left[\overline{\mathcal{M}}_{0a}^i \left(x,\bm{k}_T;S,T\right)\mathcal{M}_{0a}^j \left(x,\bm{k}_T;S,T\right)\right]\,.\label{eq:Phi}
\end{align}
The tree-level gluon-gluon correlator in this model is given by
\begin{align}
&\mathcal{M}_{0a}^j \left(x,\bm{k}_T;S,T\right)\notag\\
=&\left\langle P-k\left|F_a^{+j}\right|P,S,T\right\rangle \notag\\
=&\epsilon_c^{*\nu}\left(P-k,\lambda_S\right)G_{ab}^{j\alpha}\left(k,k\right)\mathcal{Y}_{\mu \nu \alpha,bc} \epsilon^{\mu}\left(P,\lambda\right)\,.\label{eq:amp}
\end{align}
Here, $\epsilon^{\mu}(P,\lambda)$ is the deuteron polarization vector, and $\epsilon_c^{*\nu}(P-k,\lambda_S)$ is the polarization vector of the spectator with color index $c$. The factor
\begin{align}
G_{ab}^{j\alpha}\left(p,k\right)=-\frac{ip^+}{k^2}\left(g^{j\alpha}-\frac{k^j n_-^\alpha}{p^+}\right)
\end{align}
corresponds to the Feynman rule for the field strength tensor $-i\left(p^\mu g^{\nu \rho}-p^\nu g^{\mu \rho}\right)\delta_{ab}$~\cite{Goeke:2006ef,Collins:2011zzd}. The effective vertex $\mathcal{Y}_{\mu \nu \alpha,bc}$ is modeled as
\begin{align}
\mathcal{Y}_{bc}^{\mu \nu \alpha}=&\Bigg\{\left[-g^{\mu \nu}g_1\left(k^2\right)-\frac{2 \overline{P}^\mu \overline{P}^\nu}{M^2} g_3\left(k^2\right)\right]\left(P^\alpha + \left(P-k\right)^\alpha\right)\notag\\
&+2\left(\overline{P}^\nu g^{\alpha \mu}+\overline{P}^\mu g^{\alpha \nu}\right) g_2\left(k^2\right)\Bigg\}\delta_{bc}\,,\label{eq:vertex}
\end{align}
with $\overline{P} = (P + (P-k))/2$. The functions $g_{1,2,3}(k^2)$ are the deuteron-gluon-spectator couplings. This parametrization follows the conventional form factor decomposition of the vector current for a spin-1 particle~\cite{Arnold:1979cg,Brodsky:1992px,Zhang:2024nxl}. Various functional forms for $g_i(k^2)$ exist in the literature~\cite{Bacchetta:2008af}. The simplest choice is a point-like coupling (i.e., constants $g_i$). However, this form leads to divergences upon integration over $\bm{k}_T$. To regularize these divergences, we adopt an exponential form factor,
\begin{align}
g_{1,2,3}\left(k^2\right) =\kappa_{1,2,3} e^{k^2/\Lambda_S^2}\,,
\end{align}
where $\kappa_{1,2,3}$ are free parameters and $\Lambda_S$ is a cut-off scale 

Since the spectator is on-shell, $(P-k)^2 = M_S^2$, the virtuality of the gluon is given by
\begin{align}
k^2=-\frac{\bm{k}_T^2}{1-x}-\frac{xM_S^2}{1-x}+xM^2\,.
\end{align}

Using Eqs.~(\ref{eq:Phi}) and (\ref{eq:amp}), the tree-level spectator approximation to the gluon-gluon correlator can be written as
\begin{align}
&\Phi^{ij}\left(x,\bm{k}_T;S,T\right)\notag\\
=&\frac{1}{xP^+}\frac{1}{\left(2\pi\right)^3} \frac{1}{2\left(1-x\right)P^+}\notag\\
&\times G_{a b^\prime}^{i\beta*}\left(k,k\right)G_{ab}^{j\alpha}\left(k,k\right) \mathcal{Y}^*_{\mu^\prime \nu^\prime \beta,b^\prime c^\prime}\mathcal{Y}_{\mu \nu \alpha,bc}\notag\\
&\times \epsilon_c^{*\nu}\left(P-k,\lambda_S\right) \epsilon_{c^\prime}^{\nu^\prime}\left(P-k,\lambda_S\right)\notag\\
&\times \epsilon^{\mu^\prime *}\left(P,\lambda\right)\epsilon^{\mu}\left(P,\lambda\right)\,,
\end{align}
where the summation over all polarization states of the spectator yeilds:
\begin{align}
&\sum_{\lambda_S}\epsilon_c^{*\nu}\left(P-k,\lambda_S\right) \epsilon_{c^\prime}^{\nu^\prime}\left(P-k,\lambda_S\right)\notag\\
=&-g^{\nu \nu^\prime}+\frac{\left(P-k\right)^\nu \left(P-k\right)^{\nu^\prime}}{M_S^2}\,.
\end{align}

\begin{figure*}[htbp]
    \centering
    \begin{tabular}{c@{\hspace{3em}}c}
    \includegraphics[width=0.8\columnwidth]{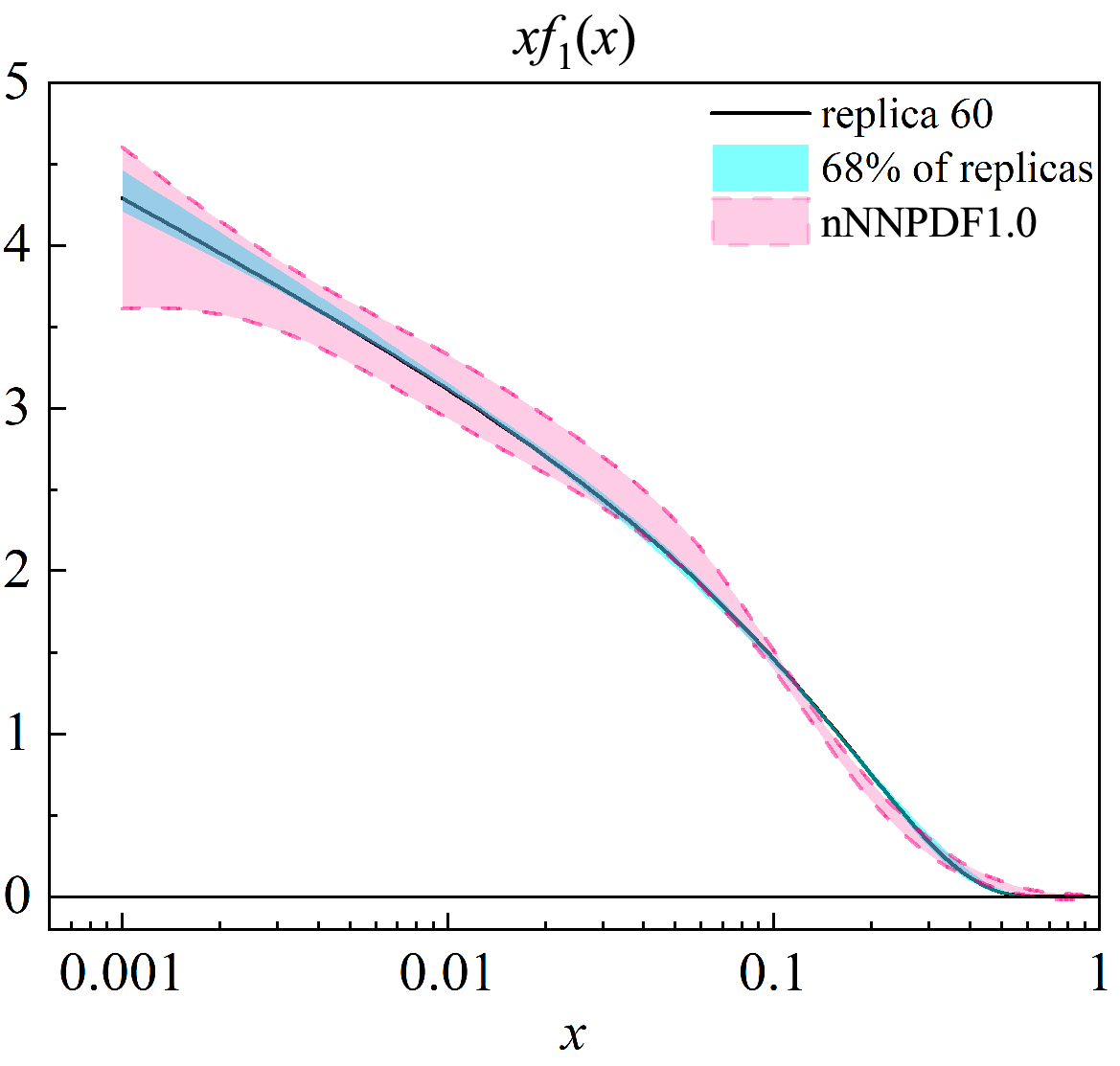}&\includegraphics[width=0.8\columnwidth]{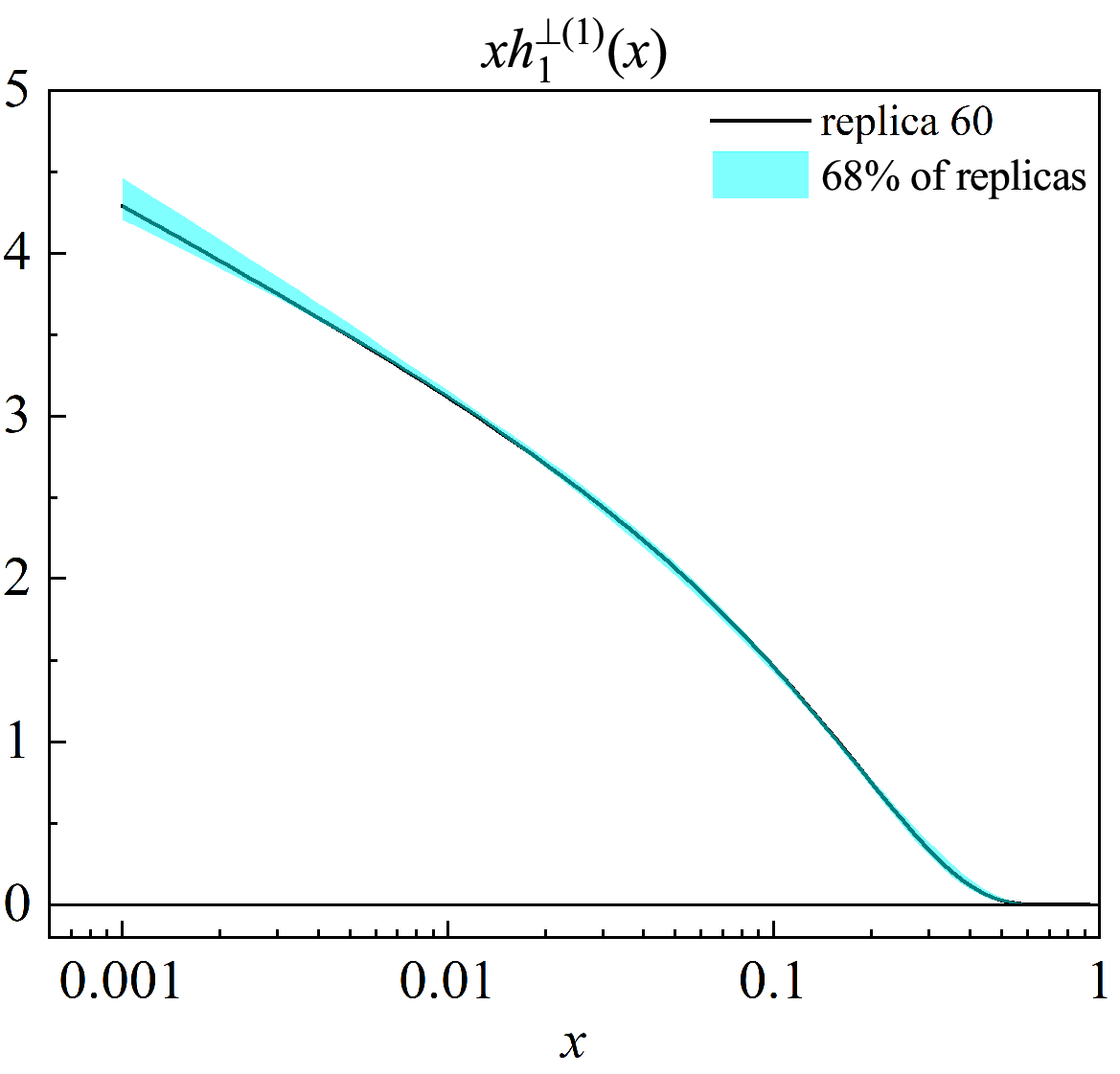}\\
    \includegraphics[width=0.8\columnwidth]{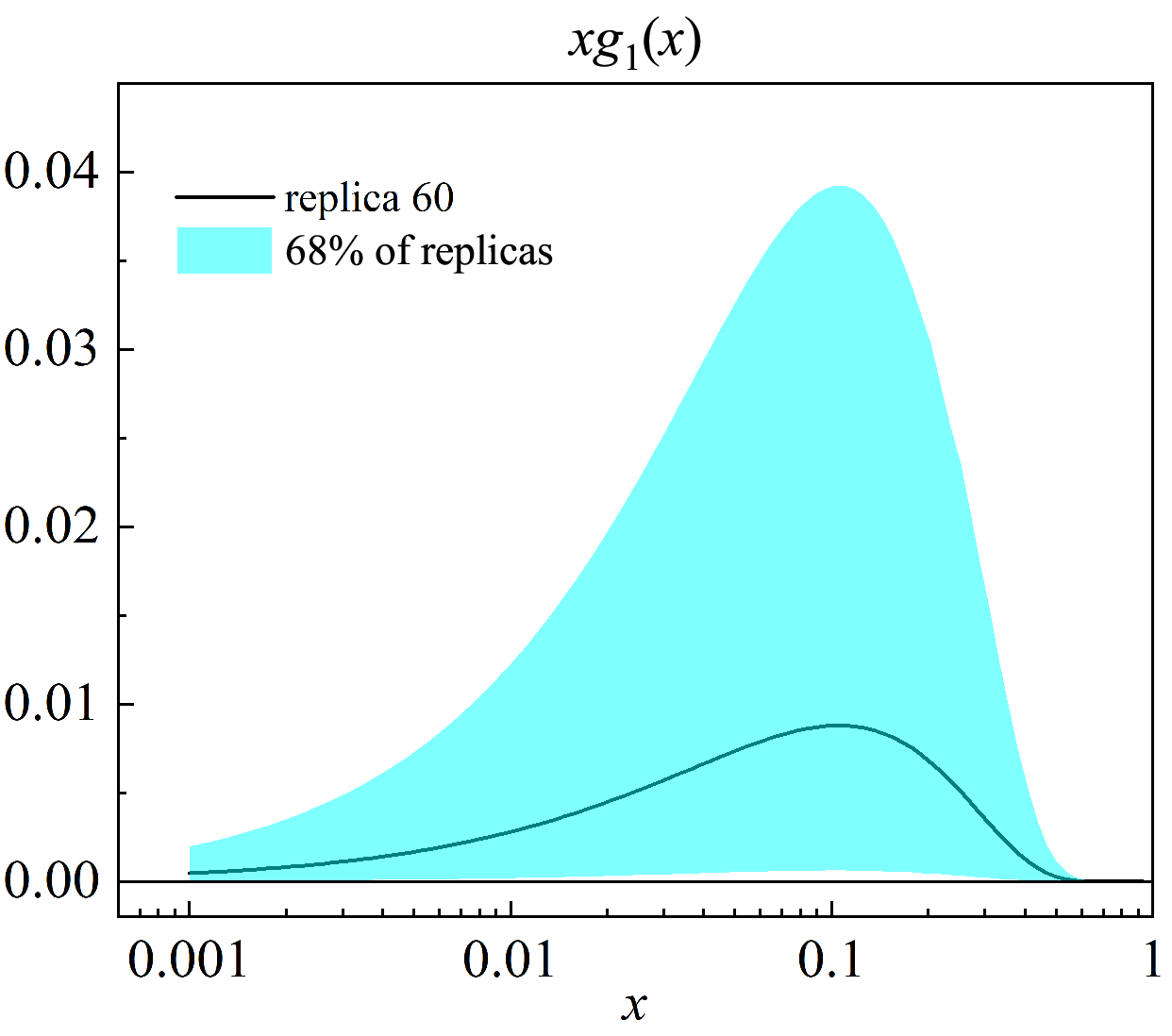}&\includegraphics[width=0.8\columnwidth]{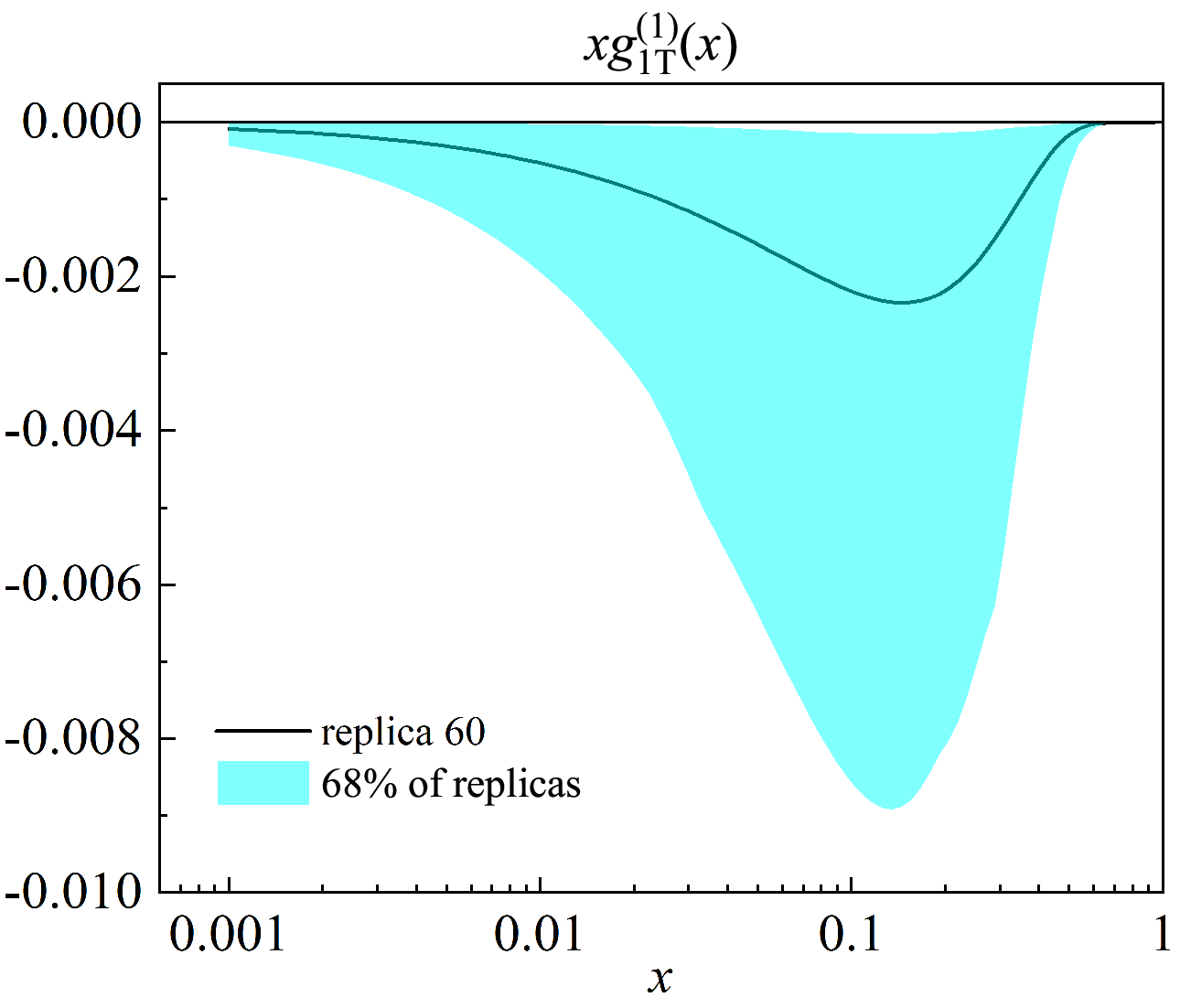}\\
    \end{tabular}
    \caption{
    Upper-left panel: Fit of the integrated unpolarized gluon TMD $xf_1(x)$ for the deuteron at $Q_0 = 2\,\mathrm{GeV}$ in the range $0.001 < x < 1$. The band with dashed borders corresponds to the nNNPDF1.0 parametrization of $xf_1$~\cite{AbdulKhalek:2019mzd}. The solid line shows the result for replica 60. The cyan band represents the 68\% uncertainty band of the spectator model fit. The other three panels display the $x$-dependence of $xh_1^{\perp(1)}$, $xg_1$, and $xg_{1T}^{(1)}$ obtained from the spectator model using the parameters in Table~\ref{table:parm}.
    }\label{fig:x1}     
\end{figure*}

\begin{figure*}[htbp]
    \centering
    \begin{tabular}{c@{\hspace{3em}}c}
    \includegraphics[width=0.8\columnwidth]{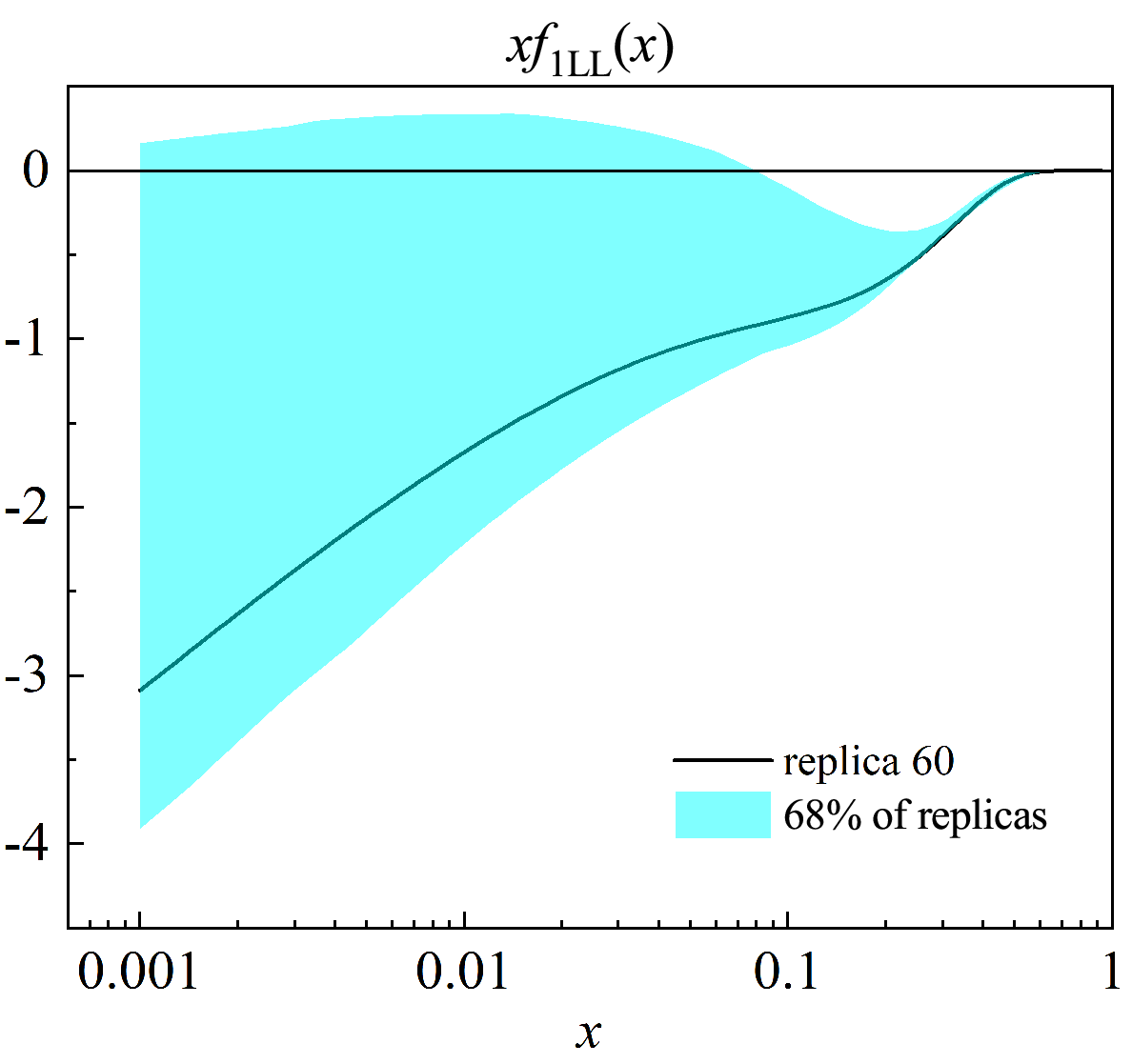}&\includegraphics[width=0.8\columnwidth]{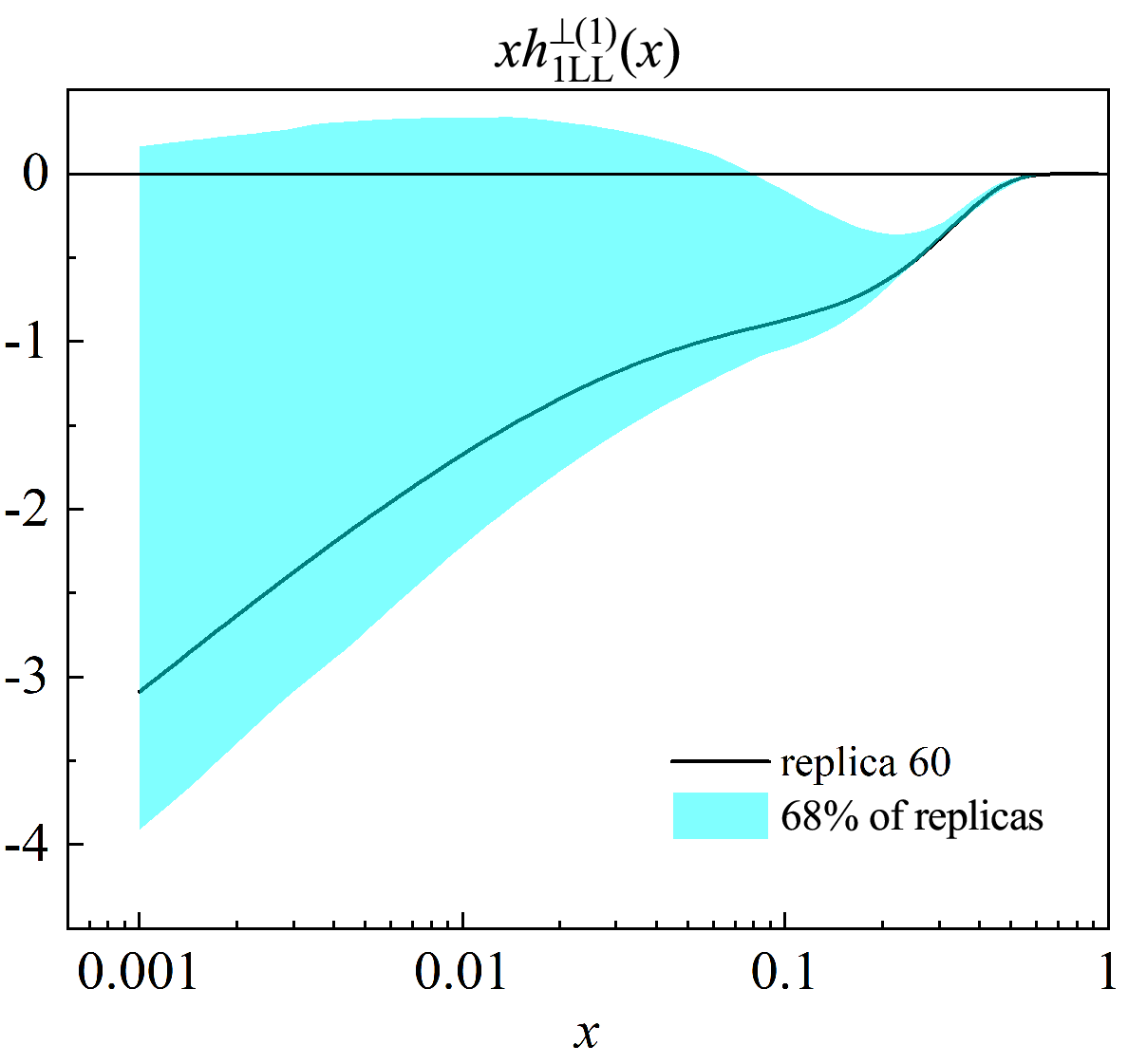}
    \end{tabular}
    \caption{
    The integrated LL tensor-polarized TMDs as functions of $x$ at $Q_0 = 2\,\mathrm{GeV}$. The band represents the 68\% uncertainty of the TMDs, and the solid line corresponds to replica 60.
    }\label{fig:x2}       
\end{figure*}

\begin{figure*}[htbp]
    \centering
    \begin{tabular}{c@{\hspace{3em}}c}
    \includegraphics[width=0.8\columnwidth]{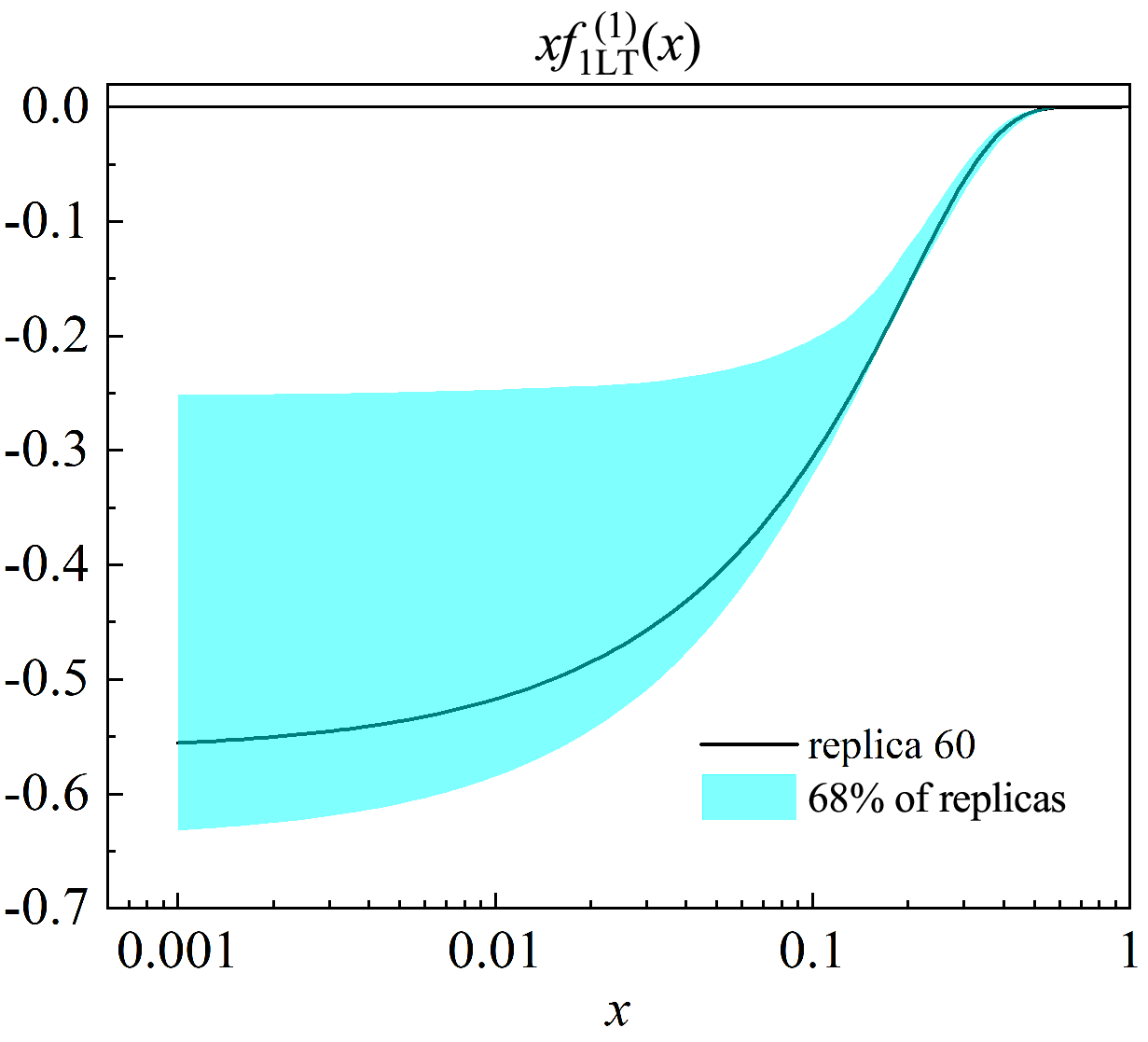}&\includegraphics[width=0.8\columnwidth]{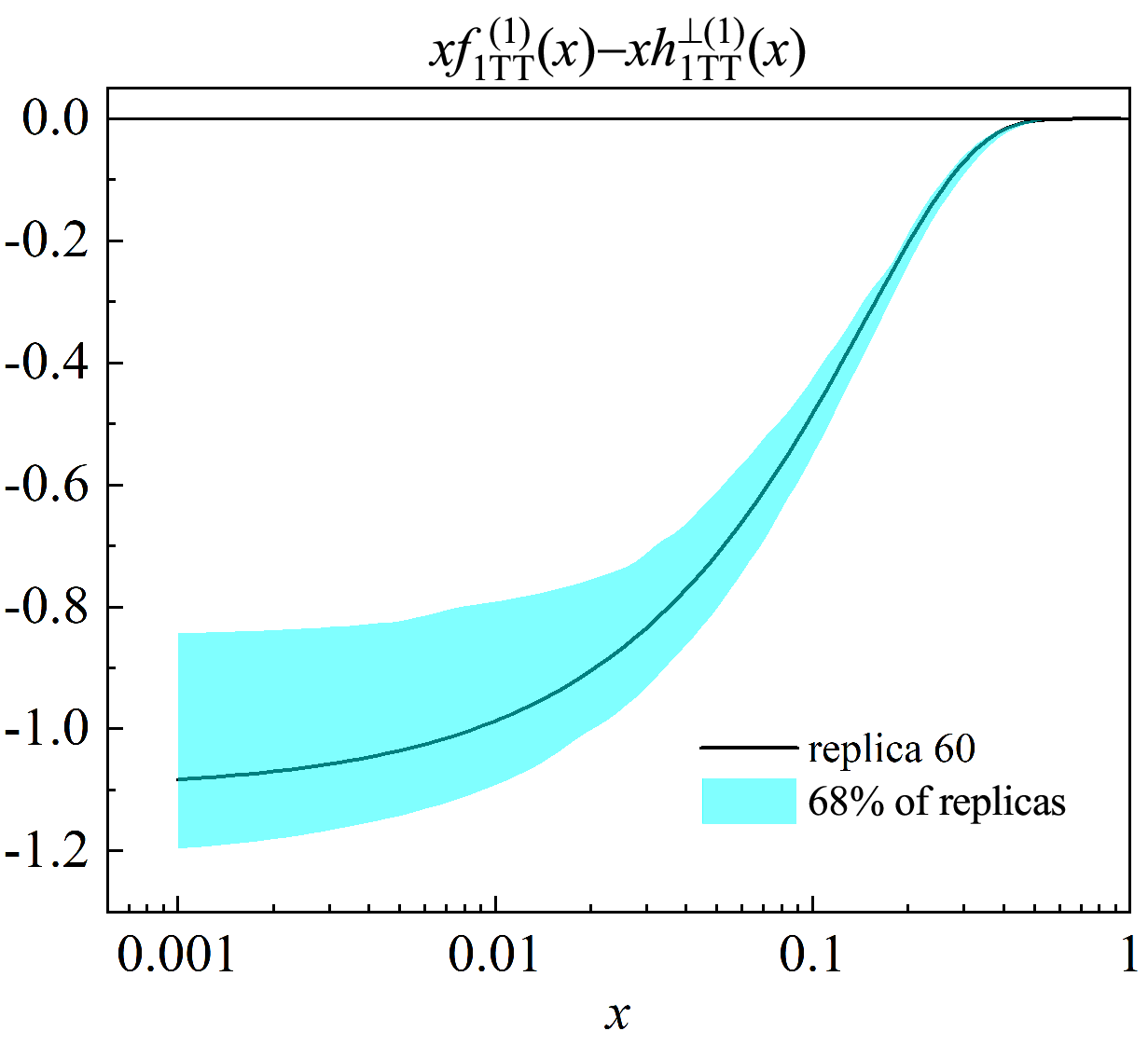}\\
    \includegraphics[width=0.8\columnwidth]{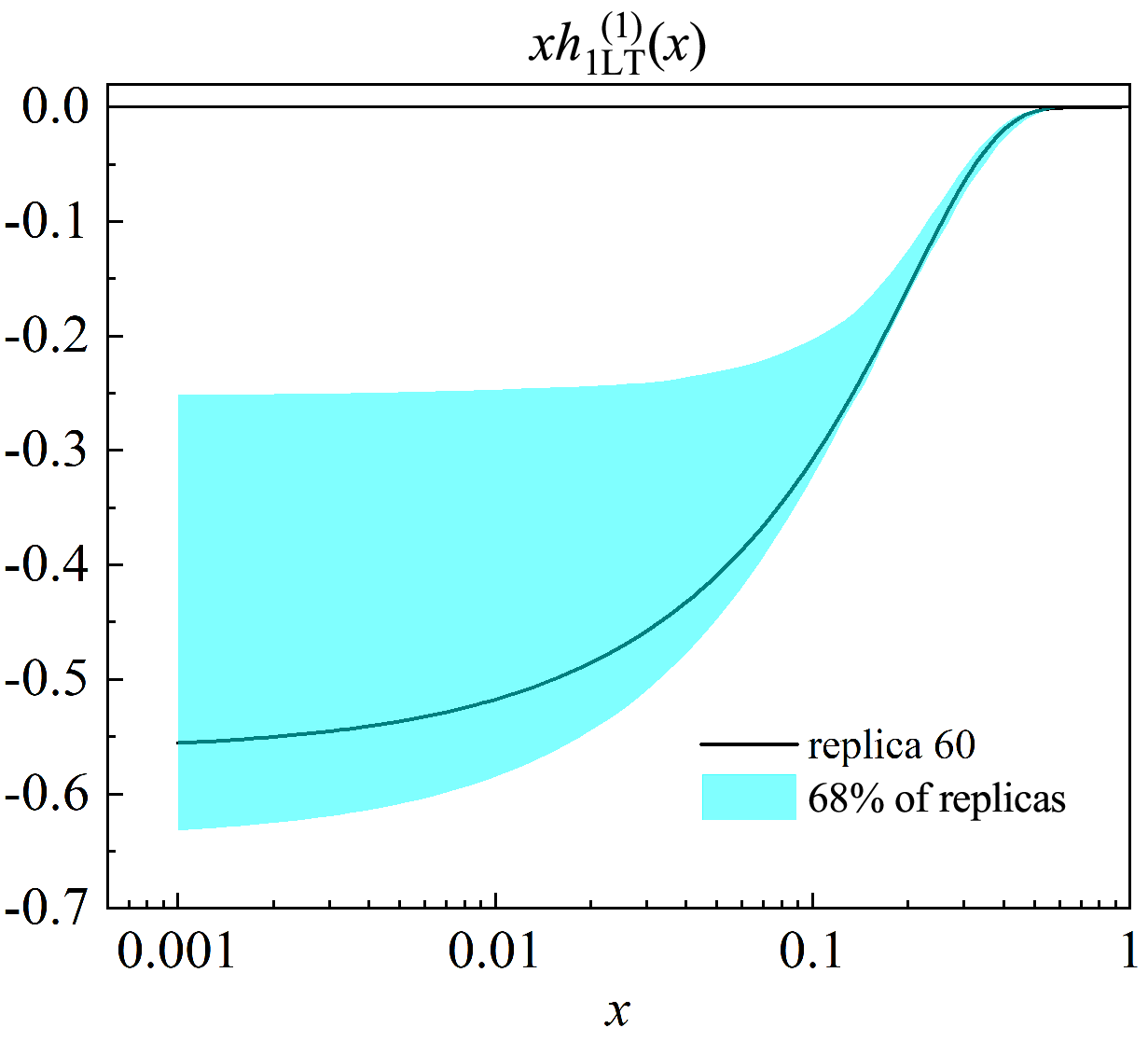}&\includegraphics[width=0.8\columnwidth]{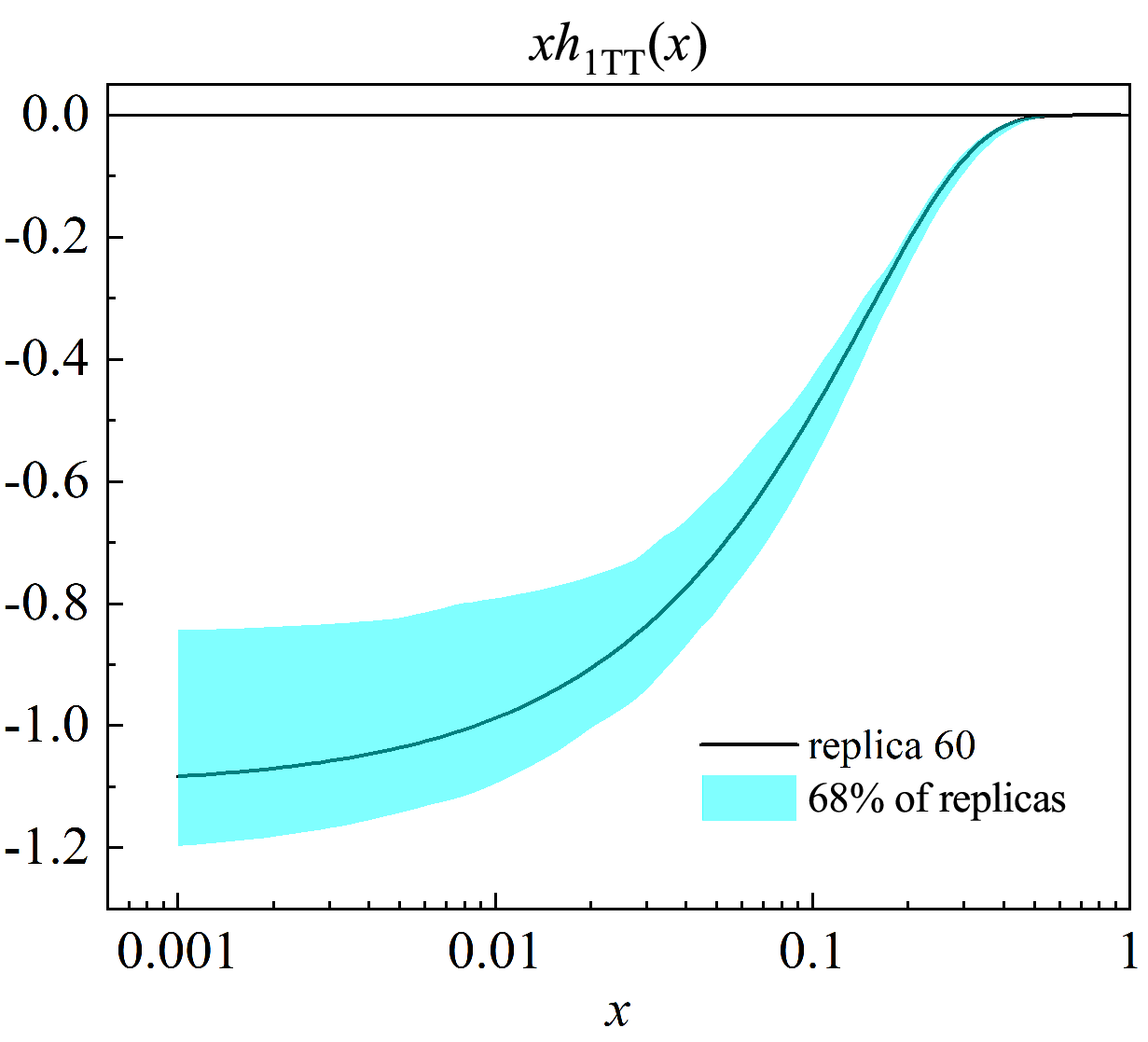}\\
    \includegraphics[width=0.8\columnwidth]{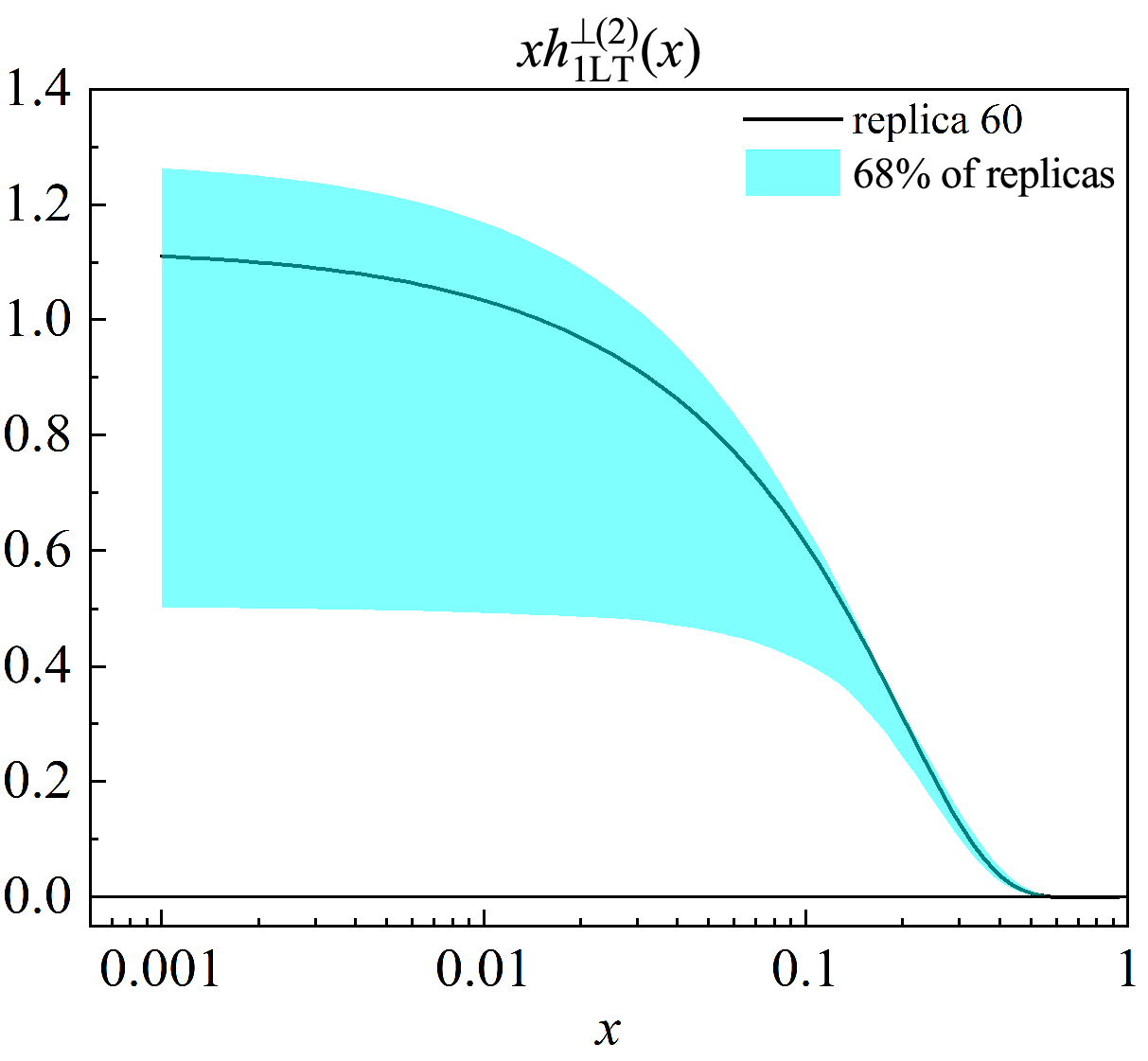}&\includegraphics[width=0.8\columnwidth]{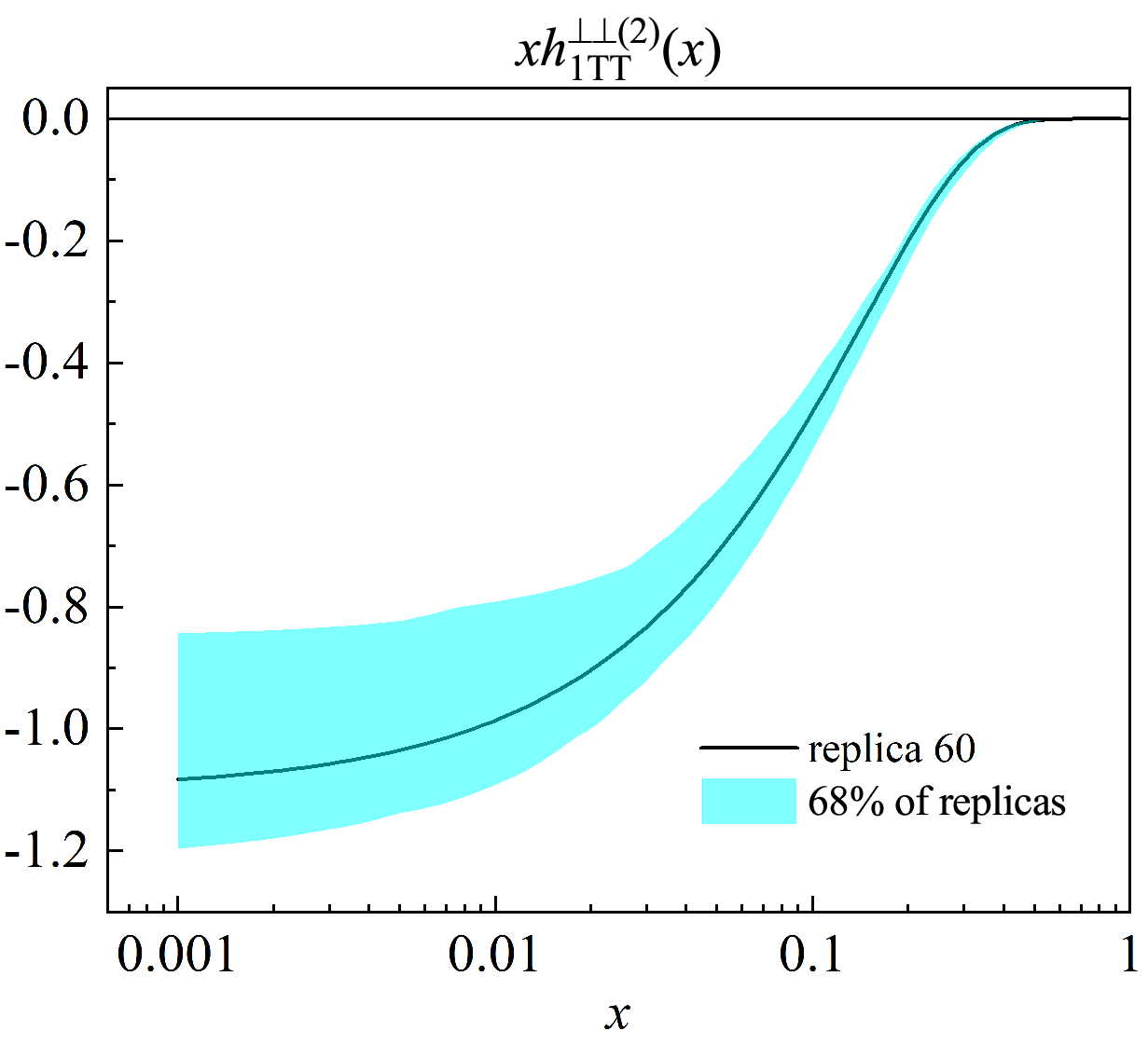}
    \end{tabular}
    \caption{
    The integrated LT and TT tensor-polarized TMDs as functions of $x$ at $Q_0 = 2\,\mathrm{GeV}$. Conventions as in Fig.~\ref{fig:x2}. Left column: LT-polarized PDFs $xf_{1LT}^{(1)}$, $xh_{1LT}^{(1)}$, and $xh_{1LT}^{\perp(2)}$. Right column: TT-polarized PDFs $xf_{1TT}^{(1)} - xh_{1TT}^{\perp(1)}$, $xh_{1TT}$, and $xh_{1TT}^{\perp\perp(2)}$.
    }\label{fig:x3}       
\end{figure*}

\begin{figure*}[htbp]
    \centering
    \includegraphics[width=1.8\columnwidth]{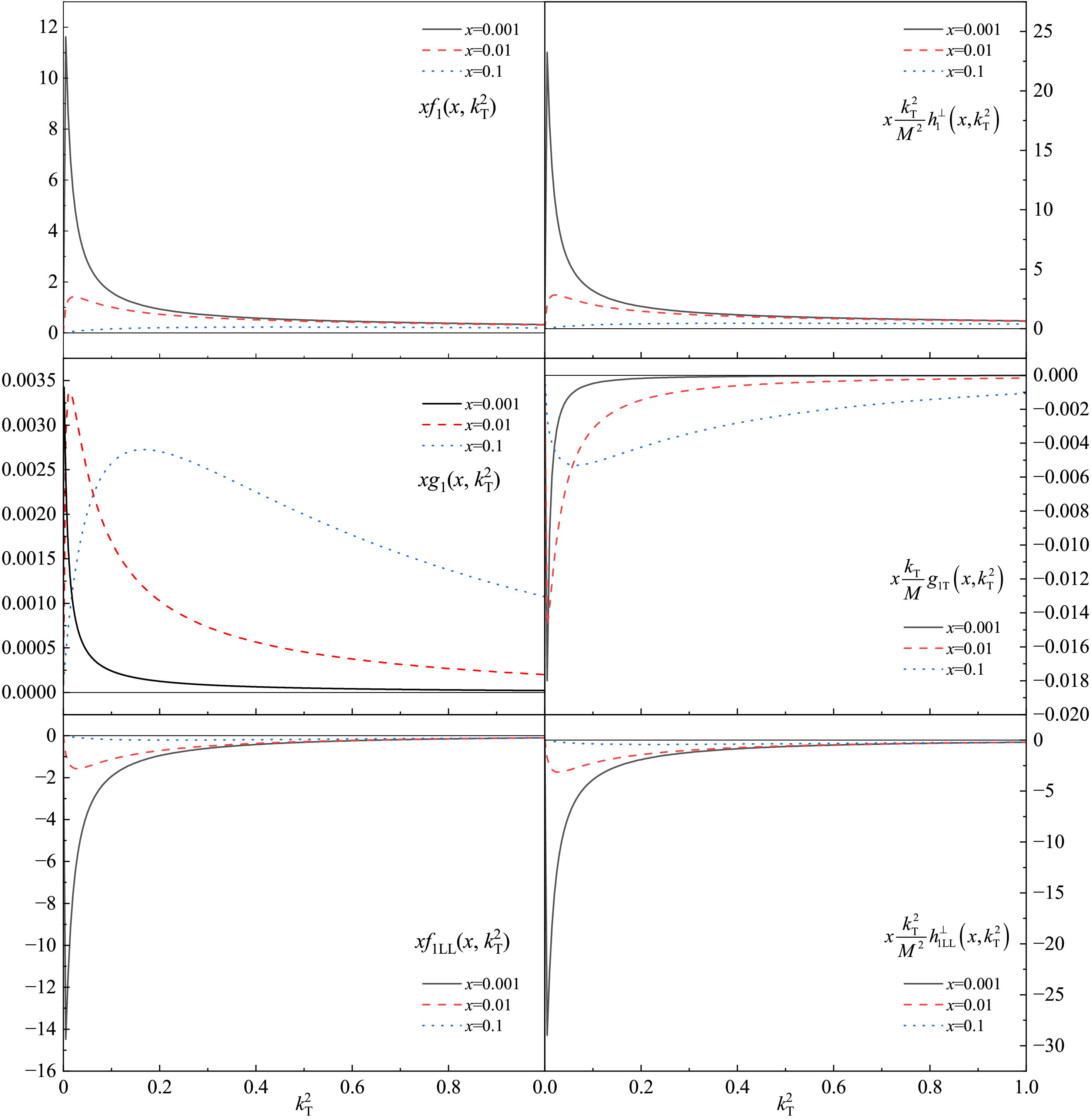}
    \caption{
    The unpolarized, vector-polarized, and LL tensor-polarized TMDs (replica 60) as functions of $\bm{k}_T^2$ for $x = 0.001$, $0.01$, and $0.1$ at $Q_0 = 2\,\mathrm{GeV}$. Solid line: $x = 0.001$; dashed line: $x = 0.01$; dotted  line: $x = 0.1$.
    }\label{fig:kt1}    
\end{figure*}
\begin{figure*}[htbp]
    \centering
    \includegraphics[width=1.8\columnwidth]{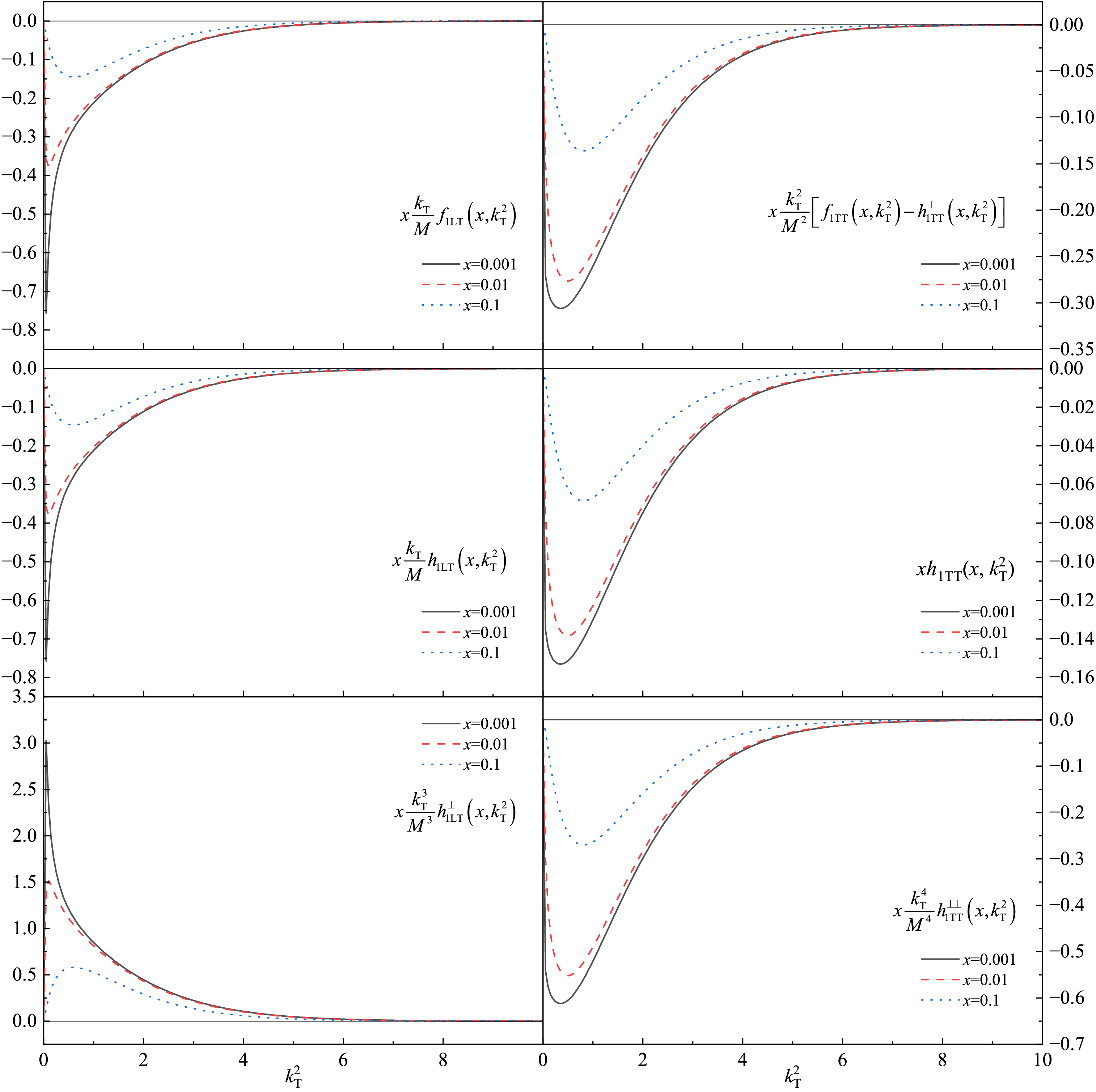}
    \caption{
    The LT and TT tensor-polarized TMDs (replica 60) as functions of $\bm{k}_T^2$ for $x = 0.001$, $0.01$, and $0.1$ at $Q_0 = 2\,\mathrm{GeV}$. Line conventions as in Fig.~\ref{fig:kt1}.
    }\label{fig:kt2}        
\end{figure*}

By projecting the correlator $\Phi^{ij}$ onto the tensors $g_T^{ij}$, $\epsilon_T^{ij}$, $k_T^i$, $k_T^{ij}$, etc., we obtain the expressions for the leading-twist T-even gluon TMDs, which are collected in Appendix~\ref{appendix1}. We observe that the coupling $g_2(k^2)$ in Eq.~\eqref{eq:vertex} contributes significantly to the final expressions. In the "$g_2$-vanishing approximation", where $g_2(k^2)$ is set to zero, the TMDs $g_1(x,\bm{k}_T^2)$ and $g_{1T}(x,\bm{k}_T^2)$ vanish, while the remaining TMDs satisfy the following relations:
\begin{align}
f_1^{(g_{1,3})}=&\frac{\bm{k}_T^2}{2M^2}h_1^{\perp(g_{1,3})}\,,\\
f_{1LL}^{(g_{1,3})}=&\frac{\bm{k}_T^2}{2M^2}h_{1LL}^{\perp(g_{1,3})}\,,\\
f_{1LT}^{(g_{1,3})}=&h_{1LT}^{(g_{1,3})}=-\frac{\bm{k}_T^2}{4M^2}h_{1LT}^{\perp(g_{1,3})}\,,\\
h_{1TT}^{(g_{1,3})}=&\frac{\bm{k}_T^2}{2M^2}\left[f_{1TT}^{(g_{1,3})}-h_{1TT}^{\perp(g_{1,3})}\right]\notag\\
=&\left(\frac{\bm{k}_T^2}{2M^2}\right)^2 h_{1TT}^{\perp \perp (g_{1,3})}\,.
\end{align}
In this approximation, the gluon TMDs saturate the positivity bounds derived in Ref.~\cite{Cotogno:2017puy}:
\begin{align}
\frac{\bm{k}_T^2}{2M^2}\left|h_1^\perp-h_{1LL}^\perp\right|\leq& f_1-f_{1LL}\,,\\
\frac{\bm{k}_T^4}{16M^4}\Big[4\left(h_{1L}^\perp\right)^2+&\big(2h_1^\perp+h_{1LL}^\perp\big)\Big]\notag\\
\leq\left(f_1+\frac{f_{1LL}}{2}+g_1\right)&\left(f_1+\frac{f_{1LL}}{2}-g_1\right)\,.
\end{align}
Furthermore, the tensor polarized TMDs $F_{(T)}^{(g_{1,3})}$ in the $g_2$-vanishing approximation exhibit the same functional dependence on the couplings $g_1$ and $g_3$:
\begin{align}
F_{(T)}^{(g_{1,3})}\sim\bigg[4M^2(1-x)g_1+\Big(\bm{k}_T^2 +(M(1-x)-M_S)^2\Big)g_3\bigg]\notag\\
\times \bigg[4M^2(1-x)g_1+\Big(\bm{k}_T^2 +(M(1-x)+M_S)^2\Big)g_3\bigg]\,.
\end{align}

Finally, in our model, the spectator mass $M_S$ is not fixed but is allowed to take a continuous range of real values, weighted by a spectral function $\rho(M_S)$:
\begin{align}
\rho\left(M_S\right)=\mu^{2a}\left[\frac{A}{B+\mu^{2b}}+\frac{C}{\pi \sigma} e^{-\frac{(M_S-D)^2}{\sigma^2}}\right]\,,
\end{align}
where $\mu^2 = M_S^2 - M^2$, and $\{A, B, a, b, C, D, \sigma\}$ are free parameters. This spectral function models the spectator mass distribution as a smooth background with a superimposed resonance peak.The gluon TMDs are then weighted by the spectral function $\rho\left(M_S\right)$:
\begin{align}
F\left(x,\bm{k}_T^2\right)=\int_{M}^{\infty}dM_S\ \rho\left(M_S\right) F\left(x,\bm{k}_T^2;M_S\right)\,,\label{eq:weighted}
\end{align}
which provides an effective way to incorporate $q\bar{q}$ contributions to the spectator system that become energetically accessible at large $M_S$.

\section{Numerical results}\label{section3}

In this section, we present the numerical results for the gluon TMDs of the tensor-polarized deuteron. Our model contains 11 free parameters: seven from the spectral function ($a$, $b$, $A$, $B$, $C$, $D$, $\sigma$) and four from the form factors ($\kappa_1$, $\kappa_2$, $\kappa_3$, $\Lambda_S$). To determine these parameters, we perform a fit to the integrated unpolarized TMD $f_1(x)$, defined as
\begin{align}
f_1\left(x\right)=\int d\bm{k}_T^2\ f_1\left(x,\bm{k}_T^2\right)\,,
\end{align}
which reproduces the collinear unpolarized gluon PDF at an initial scale. The $\bm{k}_T$-integrated TMDs are also weighted by the spectral function as in Eq.~(\ref{eq:weighted}). Additionally, we define the $n$-th moment of a TMD as
\begin{align}
F^{(n)}\left(x\right)=\int d\bm{k}_T^2 \left(\frac{\bm{k}_T^2}{2M^2}\right)^n F\left(x,\bm{k}_T^2\right)\,.
\end{align}

\begin{table}[H]
\centering
\caption{Central column:mean values and uncertainties of the fitted model parameters. Rightmost column: corresponding values for replica 60}\label{table:parm}
    \setlength{\tabcolsep}{0.4cm}{
    \begin{tabular}{ccc}
    \hline
    Parameter & Mean & Replica 60 \\
    \hline
    $\kappa_1$ &0.713$\pm$0.604 &0.350 \\
    $\kappa_2$ &0.334$\pm$0.303 &0.149 \\
    $\kappa_3$ &15.56$\pm$6.06 &12.88 \\
    $\Lambda_S$&1.34$\pm$0.18 &1.36 \\
    $a$&1.564$\pm$1.442 &1.529 \\
    $b$&10.14$\pm$5.66 &5.93 \\
    $A$&138$\pm$141 &221 \\
    $B$&5.86$\pm$6.65 &6.42 \\
    $C$&305$\pm$164 &359 \\
    $D$&1.19$\pm$0.55 &1.31 \\
    $\sigma$&0.683$\pm$0.226 &0.674 \\
    \hline
    \end{tabular}}
\end{table}

We fix our model parameters by fitting the nNNPDF1.0 parametrization of $f_1\left(x\right)$~\cite{AbdulKhalek:2019mzd} at the scale $Q_0=2~\mathrm{GeV}$. We avoid using a lower energy scale, as the parametrization in the small-$x$ region at lower scales exhibits large uncertainties, along with effects not included in our model. The fit is performed  over the range $0.001<x<1$.
To account for uncertainties in the fitting procedure, we employ the bootstrap method: we create $N=100$ replicas of the nNNPDF1.0 central value, each modified by adding random Gaussian noise with the same variance as the uncertainty of the original parametrization. 
For each replica, we perform a separate fit to obtain $N$ parameter sets. The 68\% uncertainty band of our fit is constructed by rejecting the largest and smallest 16\% of the $N$ values of any prediction.

The values of the model parameters are shown in Tab.~\ref{table:parm}. 
The second column lists the parameter values and their uncertainties, calculated as the average and the uncertainty given by the semi-difference between the upper and lower limits of the distribution. We obtain a total $\chi^2/\text{d.o.f.} = 1.75 \pm 0.28$. 
The final column, labeled ``Replica 60", shows the most representative parameter set, which has the minimal deviation from the mean values.

In Fig.~\ref{fig:x1}, we show the numerical results of our fit for $x f_1(x)$ at $Q_0 = 2\ \textrm{GeV}$ and the integrated vector polarized gluon TMDs as functions of $x$. The results indicate that the effects of these TMDs are significant and could be probed by future experimental measurements. The band with dashed borders represents the nNNPDF1.0 parametrization result. The solid line corresponds to the fit from replica 60, and the cyan band shows the 68\% uncertainty of the spectator model fit. The size of $xg_1\left(x\right)$ and $xg_{1T}^{(1)}\left(x\right)$ is several times smaller than that of $xf_1\left(x\right)$. 
As discussed in Section.~\ref{section2}, this indicates that the coupling $g_2\left(k^2\right)$ in Eq.~(\ref{eq:vertex})  has a small but necessary impact on the numerical results of gluon TMDs for the deuteron. The deuteron momentum and spin fractions carried by gluons at the scale $Q_0=2\mathrm{GeV}$ in our model are predicted as:

\begin{align}
\left\langle x\right\rangle_g=&\int_{0}^{1} dx ~ xf_1\left(x\right)=0.412\pm 0.009\,,\\
\left\langle 1\right\rangle_g=&\int_{0}^{1} dx ~ g_1\left(x\right)=0.064\pm 0.056\,,
\end{align}
We anticipate future parametrizations of $g_1$ that enable simultaneous fitting of both $f_1$ and $g_1$, thereby improving the predictive power of the model. The notable dependence of the TMDs $g_1\left(x\right)$ and $g_{1T}^{(1)}\left(x\right)$ on the coupling $g_2\left(k^2\right)$ underscores the need for more comprehensive future studies.

In Figs.~\ref{fig:x2} and \ref{fig:x3}, we display the integrated tensor-polarized gluon TMDs as functions of $x$ at the scale $Q_0=2~\mathrm{GeV}$, without including the evolution effects. 
For the deuteron with the same polarization, the gluon TMDs exhibit similar tendency. The TMDs $xh_{1}^{\perp(1)}\left(x\right)$, $xg_{1}\left(x\right)$ and $xh_{1LT}^{\perp(2)}$ are positive, while the remaining TMDs are negative in the entire $x$ region. 
Upon integrating the TMD correlator in Eq.~(\ref{eq:phi0}) over transverse momentum, only $f_{1LL}\left(x\right)$ and $h_{1TT}\left(x\right)$ do not vanish. 
The upper boundary of the $xf_{1LL}\left(x\right)$ band follows a tendency analogous to that of the deuteron structure $b_1\left(x\right)$ (or $xf_{1LL}\left(x\right)$ for quarks)~\cite{HERMES:2005pon}. Another integral observable is the gluon transversity $\Delta_T g\left(x\right)$ ($\equiv -h_{1TT}\left(x\right)$), which does not exist in the spin-1/2 nucleons, as a change of two spin units is required for the helicity-flip amplitude. In our model, $xh_{1TT}\left(x\right)$ increases with $x$.

In Figs.~\ref{fig:kt1} and \ref{fig:kt2}, the T-even gluon TMDs multiplied by $x\bm{k}_T^n/M^n$ are displayed as functions of $\bm{k}_T^2$ at $x=10^{-3}$ (solid), $x=10^{-2}$ (dashed) and $x=10^{-1}$ (dotted), at the scale $Q_0=2~\mathrm{GeV}$. The parameters are taken from the most representative replica 60. The absolute magnitudes of $xg_1 \left(x,\bm{k}_T^2\right)$ and $x \bm{k}_T /M~g_{1T}\left(x,\bm{k}_T^2\right)$ are smaller. These TMDs share a common trend: they exhibit a non-Gaussian shape in $\bm{k}_T^2$, with a long tail as $\bm{k}_T^2$ increases. They are predominantly concentrated in the small $\bm{k}_T^2$ region and display a peak, that shifts to larger $\bm{k}_T^2$ and becomes flatter as $x$ increases. Notably, $xf_1\left(x,\bm{k}_T^2\right)$ do not vanish at $\bm{k}_T^2 \rightarrow 0$ GeV$^2$, indicating significant contribution from gluon orbital angular momentum $L=1$.

Finally, we verify that our model results satisfy the positivity bounds outlined in Appendix~\ref{appendix2}, which provide important model-independent constraints for both TMD and collinear cases~\cite{Cotogno:2017puy}, while neglecting T-odd gluon TMDs.

\section{Conclusions}\label{section4}
In this paper, we presented a model calculation of leading-twist T-even gluon TMDs for the tensor-polarized deuteron within a spectator model framework. The model is based on the assumption that an on-shell deuteron can emit a time-like off-shell gluon, with the remaining system is treated as a single on-shell spectator particle whose mass can take real values within a continuous range, described by a spectral function. For spin-1 hadrons, the polarization description requires not only a spin vector $S$ but also a symmetric traceless spin tensor $T$ to characterize tensor polarized states. The deuteron-gluon-spectator vertex is modeled to the vector current for a spin-1 particle, and we adopted an exponential form factor for the deuteron-gluon-spectator couplings. 

We derived analytical expressions for thirteen T-even gluon TMDs and provided numerical results for their $x$-dependence and $\bm{k}_T$-dependence. In the fit procedure, we employed the bootstrap method to account for the uncertainties, fixing the model parameters by fitting the nNNPDF1.0 parametrization for $f_1\left(x\right)$ at the scale $Q_0=2~\mathrm{GeV}$. We also verified that our results satisfy the positivity bounds for both the gluon TMDs and their integrated counterparts. Our analysis reveals non-negligible impact of these gluon TMDs, particularly for tensor polarized hadrons, which could be explored in future experimental measurements. 

As is well known, gluon TMDs exhibit a more intricate dependence on the structure of the color flow (gauge link), which in turn introduces a dependence on the involved process. There are two main classes of gluon TMDs: the so-called Weizs$\ddot{a}$cker-Williams (WW) gluon TMDs ($f$-type) and the dipole gluon TMDs ($d$-type). In future work, we plan to extend our model to include leading-twist T-odd gluon TMDs for the tensor polarized deuteron. It would be particularly interesting to investigate the differences between $f$-type and $d$-type gluon TMDs within this framework.

\section*{Acknowledgements}
This work is partially supported by the National Natural Science Foundation of China under Grant Nos. 12150013 and 12335001, as well as supported, in part, by National Key Research and Development Program under the contract No. 2024YFA1610503. Xiupeng Xie is also supported by the SEU Innovation Capability Enhancement Plan for Doctoral Students under Grant number CXJH$\_$SEU 25138.

\begin{widetext}
\appendix
\section{Full expressions of gluon TMDs}\label{appendix1}
In the following, we list the final expressions of the gluon TMDs in Eqs.~(\ref{eq:phiU})-(\ref{eq:phiTT}). The unpolarized gluon TMD $f_1$ is given by
\begin{align}
f_1\left(x,\bm{k}_T^2\right)=&-g_T^{ij} \Phi_U^{ij}\left(x,\bm{k}_T\right) \notag\\
=&-\Bigg\{-8 g_3 M^2 \left(x-1\right) \bm{k}_T^2 \bigg[2 \bm{k}_T^2 \left(M^2(x-1)^2+M_S^2\right)+\bm{k}_T^4+\left(M_S^2-M^2(x-1)^2\right){}^2\bigg] \bigg[g_1 \left(\bm{k}_T^2+M^2(x-1)^2\right.\notag\\
&\left.+M_S^2\right)-g_2 \left(\bm{k}_T^2+x \left(M^2(x-1)+M_S^2\right)\right)\bigg]+g_3^2 \bm{k}_T^2\bigg[2 \bm{k}_T^2\left(M^2 (x-1)^2+M_S^2\right)+\bm{k}_T^4+\left(M_S^2 \right.\notag\\
&\left.-M^2(x-1)^2\right)^2\bigg]^2+8 M^4 (x-1)^2 \bigg[-4 g_1 g_2 \bm{k}_T^2\Big(\bm{k}_T^2+M^2 (x-1)^2+M_S^2\Big) \Big(\bm{k}_T^2+x \left(M^2(x-1)+M_S^2\right)\Big)\notag\\
&+2 g_1^2 \bm{k}_T^2 \Big(2 M_S^2\left(\bm{k}_T^2+5 M^2 (x-1)^2\right)+\left(\bm{k}_T^2+M^2(x-1)^2\right)^2+M_S^4\Big)+g_2^2 \Big(\bm{k}_T^4 \left(M^2 (x (5x-6)+2)\right.\notag\\
&\left.+(x (x+2)+2) M_S^2\right)+2 x \bm{k}_T^2 \left(M^4 (x-1)^2 (2x-1)+M^2 x ((x-1) x+1) M_S^2+(x+1) M_S^4\right)+2\bm{k}_T^6\notag\\
&+\left(M^2+M_S^2\right) \left(M^2 (x-1)^2 x-xM_S^2\right){}^2\Big)\bigg]\Bigg\}\notag\\
&\times\Bigg\{768 \pi ^3 M^6 (x-1)^3 x M_S^2\left(\bm{k}_T^2+x \left(M^2 (x-1)+M_S^2\right)\right)^2\Bigg\}^{-1}\,.
\end{align}

The vector polarized gluon TMDs are given by
\begin{align}
h_1^\perp\left(x,\bm{k}_T^2\right)=&\frac{4M^2 k_T^{ij}}{\bm{k}_T^4} \Phi_U^{ij}\left(x,\bm{k}_T\right) \notag\\
=&-\Bigg\{-8 g_3 M^2 (x-1) \bigg[2 \bm{k}_T^2 \left(M^2(x-1)^2+M_S^2\right)+\bm{k}_T^4+\left(M_S^2-M^2(x-1)^2\right)^2\bigg]\notag\\
&\times \bigg[g_1 \left(\bm{k}_T^2+M^2(x-1)^2+M_S^2\right)-g_2 \left(\bm{k}_T^2+x \left(M^2(x-1)+M_S^2\right)\right)\bigg]+g_3^2 \Big(2 \bm{k}_T^2  \left(M^2(x-1)^2+M_S^2\right)\notag\\
&+\bm{k}_T^4+\left(M_S^2-M^2(x-1)^2\right)^2\Big)^2+16 M^4 (x-1)^2 \bigg[g_1^2 \Big(2M_S^2 \left(\bm{k}_T^2+5 M^2 (x-1)^2\right)\notag\\
&+\big(\bm{k}_T^2+M^2(x-1)^2\big)^2+M_S^4\Big)-2 g_2 g_1 \left(\bm{k}_T^2+M^2(x-1)^2+M_S^2\right) \left(\bm{k}_T^2+x \left(M^2(x-1)+M_S^2\right)\right)\notag\\
&+g_2^2 \big(\bm{k}_T^2+M^2(x-1)^2+M_S^2\big) \left(\bm{k}_T^2+x \left(M^2(x-1)+M_S^2\right)\right)\bigg]\Bigg\}\notag\\
&\times\Bigg\{384 \pi ^3 M^4 (x-1)^3 x M_S^2\left(\bm{k}_T^2+x \left(M^2 (x-1)+M_S^2\right)\right)^2\Bigg\}^{-1}\,,\\
g_1\left(x,\bm{k}_T^2\right)=&-\frac{i \epsilon_T^{ij}}{S_L} \Phi_L^{ij}\left(x,\bm{k}_T\right)\notag\\
=&-g_2 \Bigg\{g_2 M^2 \bigg[2 \bm{k}_T^2 \Big(M^2 (2 x-1) (x-1)^2+(2(x-1) x+1) M_S^2\Big)+(3 x-2) \bm{k}_T^4+x \left(M_S^2-M^2(x-1)^2\right)^2\bigg]\notag\\
&+\bm{k}_T^2 \bigg[g_3 \Big(2 M_S^2\left(\bm{k}_T^2-M^2 (x-1)^2\right)+\left(\bm{k}_T^2+M^2(x-1)^2\right)^2+M_S^4\Big)-4 g_1 M^2 (x-1)\notag\\
&\times \left(\bm{k}_T^2+M^2(x-1)^2+(3-2 x) M_S^2\right)\bigg]\Bigg\}\notag\\
&\times\Bigg\{64 \pi ^3 M^2 (x-1) M_S^2 \left(\bm{k}_T^2+x \left(M^2 (x-1)+M_S^2\right)\right)^2\Bigg\}^{-1}\,,\\
g_{1T}\left(x,\bm{k}_T^2\right)=&-\frac{i \epsilon_T^{ij}M}{\bm{k}_T \cdot \bm{S}_T} \Phi_T^{ij}\left(x,\bm{k}_T\right)\notag\\
=&-g_2 \Bigg\{4 g_1 M^2 (x-1) \left(-\bm{k}_T^2+M^2 (x-1)^2-M_S^2\right)\left(\bm{k}_T^2+M^2 (x-1)^2+(3-2 x) M_S^2\right)+4 g_2 M^2 (x-1)\notag\\
&\times\bigg[M_S^2 \left((x+1) \bm{k}_T^2-M^2 (x-1)^2 x\right)+M^2 (x-1)^2\bm{k}_T^2+\bm{k}_T^4+x M_S^4\bigg]-g_3 \left(-\bm{k}_T^2+M^2(x-1)^2-M_S^2\right)\notag\\
&\times \left(\bm{k}_T^2+\left(M_S+M(x-1)\right)^2\right)\left(\bm{k}_T^2+\left(M_S+M(-x)+M\right)^2\right)\Bigg\}\notag\\
&\times \Bigg\{128 \pi ^3 M^2 (x-1)^2 M_S^2\left(\bm{k}_T^2+x \left(M^2 (x-1)+M_S^2\right)\right)^2\Bigg\}^{-1}\,.
\end{align}

Then, the LL tensor polarized gluon TMDs are given by
\begin{align}
f_{1LL}\left(x,\bm{k}_T^2\right)=&\frac{-g_T^{ij}}{S_{LL}}\Phi_{LL}^{ij}\left(x,\bm{k}_T\right)\notag\\
=&\Bigg\{\bm{k}_T^2 \bigg[2 \bm{k}_T^2 \left(M_S^2-2 M^2(x-1)^2\right)+\bm{k}_T^4+\big(M_S^2-M^2 (x-1)^2\big)^2\bigg]\bigg[g_3 \left(\bm{k}_T^2+\left(M_S+M (x-1)\right)^2\right)\notag\\
&-4 g_1M^2 (x-1)\bigg] \bigg[g_3 \left(\bm{k}_T^2+\left(M_S+M(-x)+M\right)^2\right)-4 g_1 M^2 (x-1)\bigg]-4 g_2^2 M^4(x-1)^2\notag\\
&\times \bigg[\bm{k}_T^4 \Big(M^2 (x (5 x-6)+2)-2 (x (x+2)+2)M_S^2\Big)+2 x \bm{k}_T^2 \Big(M^4 (x-1)^2 (2 x-1)\notag\\
&+M^2 (2 x (2 (x-1)x-1)+3) M_S^2-2 (x+1) M_S^4\Big)-4 \bm{k}_T^6+\left(M^2-2M_S^2\right) \left(M^2 (x-1)^2 x-x M_S^2\right){}^2\bigg]\notag\\
&-4 g_2M^2 (x-1) \bm{k}_T^2 \bigg[g_3 \Big(\bm{k}_T^4 \left(M^2 (x-1) (3 x-2)-2(x+2) M_S^2\right)+2 \bm{k}_T^2 \big(M^4 (x-1)^3 (3 x-2)\notag\\
&+M^2 (1-2 x)^2(x-1) M_S^2-(2 x+1) M_S^4\big)-2 \bm{k}_T^6+x \left(M^2 (x-1)-2M_S^2\right) \left(M_S^2-M^2 (x-1)^2\right)^2\Big)\notag\\
&-4 g_1 M^2(x-1) \Big(\bm{k}_T^2 \big(M^2 (x-1) (5 x-4)-2 (x+1) M_S^2\big)-2\bm{k}_T^4\notag\\
&+x \Big(M^4 (x-1)^3+M^2 (x-1) (4 x-3) M_S^2-2M_S^4\Big)\Big)\bigg]\Bigg\}\notag\\
&\times\Bigg\{384 \pi ^3 M^6 (x-1)^3 xM_S^2\left(\bm{k}_T^2+x \left(M^2 (x-1)+M_S^2\right)\right)^2\Bigg\}^{-1}\,,\\
h_{1LL}^\perp\left(x,\bm{k}_T^2\right)=&\frac{4M^2 k_T^{ij}}{\bm{k}_T^4 S_{LL}}\Phi_{LL}^{ij}\left(x,\bm{k}_T\right)\notag\\
=&\Bigg\{g_3^2 \Big(\bm{k}_T^2+\left(M_S+M (x-1)\right)^2\Big)\Big(\bm{k}_T^2+\big(M_S-xM+M\big)^2\Big) \Big(2 \bm{k}_T^2\left(M_S^2-2 M^2 (x-1)^2\right)+\bm{k}_T^4\notag\\
&+\big(M_S^2-M^2(x-1)^2\big)^2\Big)+16 g_1^2 M^4 (x-1)^2 \bigg[2 \bm{k}_T^2\big(M_S^2-2 M^2 (x-1)^2\big)+\bm{k}_T^4+\left(M_S^2-M^2(x-1)^2\right)^2\bigg]\notag\\
&-8 g_2^2 M^4 (x-1)^2 \bigg[\bm{k}_T^2\Big(M^2 (x-1) (2 x-1)-2 (x+1) M_S^2\Big)-2 \bm{k}_T^4+x \Big(M^4(x-1)^3\notag\\
&+M^2 (x-1) (4 x-3) M_S^2-2 M_S^4\Big)\bigg]-4 g_2 g_3M^2 (x-1) \bigg[\bm{k}_T^4 \Big(M^2 (x-1) (3 x-2)-2 (x+2)M_S^2\Big)\notag\\
&+2 \bm{k}_T^2 \Big(M^4 (x-1)^3 (3 x-2)+M^2 (1-2 x)^2 (x-1)M_S^2-(2 x+1) M_S^4\Big)-2 \bm{k}_T^6\notag\\
&+x \left(M^2 (x-1)-2M_S^2\right) \left(M_S^2-M^2 (x-1)^2\right)^2\bigg]-8 g_1 M^2(x-1) \bigg[g_3 \left(\bm{k}_T^2+M^2 (x-1)^2+M_S^2\right)\notag\\
&\times \Big(2 \bm{k}_T^2\left(M_S^2-2 M^2 (x-1)^2\right)+\bm{k}_T^4+\big(M_S^2-M^2(x-1)^2\big)^2\Big)-2 g_2 M^2 (x-1)\notag\\
&\times \Big(\bm{k}_T^2 \big(M^2(x-1) (5 x-4)-2 (x+1) M_S^2\big)-2 \bm{k}_T^4+x \big(M^4 (x-1)^3\notag\\
&+M^2(x-1) (4 x-3) M_S^2-2 M_S^4\big)\Big)\bigg]\Bigg\}\notag\\
&\times\Bigg\{192 \pi ^3 M^4(x-1)^3 x M_S^2 \left(\bm{k}_T^2+x \left(M^2(x-1)+M_S^2\right)\right)^2\Bigg\}^{-1}\,.
\end{align}

The LT tensor polarized gluon TMDs are given by
\begin{align}
f_{1LT}\left(x,\bm{k}_T^2\right)=&\frac{M}{2\left(\left(\bm{k}_T \cdot \bm{S}_{LT}\right)^2-\bm{k}_T^2 \bm{S}_{LT}^2 \right)}\Bigg(S_{LT}^{\left\{i\right.} \bm{k}_T^{\left.j\right\}} + \frac{-3\left(\bm{k}_T \cdot \bm{S}_{LT}\right)^2+3\bm{k}_T^2 \bm{S}_{LT}^2 }{\bm{k}_T \cdot \bm{S}_{LT}}g_T^{ij}\notag\\
&+\frac{-4\left(\bm{k}_T \cdot \bm{S}_{LT}\right)^2+2\bm{k}_T^2 \bm{S}_{LT}^2}{\bm{k}_T \cdot \bm{S}_{LT}}\frac{\bm{k}_T^i \bm{k}_T^j}{\bm{k}_T^2}\Bigg)\Phi_{LT}^{ij}\left(x,\bm{k}_T\right)\notag\\
=&-\Bigg\{8 M^4 (x-1)^2 \bigg[g_2^2 \Big(x M_S^2 \big((2 x+1) \bm{k}_T^2-2M^2 (x-1)^2 x\big)+\bm{k}_T^4+2 x^2 M_S^4\Big)+g_1 g_2 \bm{k}_T^2\Big(-3 \bm{k}_T^2\notag\\
&+M^2 (x-1)^2-(2 x+1) M_S^2\Big)+2 g_1^2 \bm{k}_T^2\Big(\bm{k}_T^2-M^2 (x-1)^2+M_S^2\Big)\bigg]-2 g_3 M^2 (x-1) \bm{k}_T^2\notag\\
&\times \bigg[4 g_1 \Big(\left(\bm{k}_T^2+M_S^2\right)^2-M^4(x-1)^4\Big)+g_2 \Big(-2 \bm{k}_T^2 \left(M^2 (x-1)^2+(2 x+1)M_S^2\right)-3 \bm{k}_T^4\notag\\
&+M^4 (x-1)^4+2 M^2 (2 x-1) (x-1)^2 M_S^2+(1-4x) M_S^4\Big)\bigg]+g_3^2 \bigg[\bm{k}_T^6 \left(M^2 (x-1)^2+3M_S^2\right)\notag\\
&-\bm{k}_T^2 \left(M^2 (x-1)^2-M_S^2\right)^3+\bm{k}_T^4\Big(-M^4 (x-1)^4-2 M^2 (x-1)^2 M_S^2+3M_S^4\Big)+\bm{k}_T^8\bigg]\Bigg\}\notag\\
&\times\Bigg\{256 \pi ^3 M^4 (x-1)^2 x M_S^2\left(\bm{k}_T^2+x \left(M^2 (x-1)+M_S^2\right)\right)^2\Bigg\}^{-1}\,,\\
h_{1LT}\left(x,\bm{k}_T^2\right)=&\frac{M}{2\left(\left(\bm{k}_T \cdot \bm{S}_{LT}\right)^2-\bm{k}_T^2 \bm{S}_{LT}^2 \right)}\Bigg(-S_{LT}^{\left\{i\right.} \bm{k}_T^{\left.j\right\}} + \frac{\left(\bm{k}_T \cdot \bm{S}_{LT}\right)^2-\bm{k}_T^2 \bm{S}_{LT}^2 }{\bm{k}_T \cdot \bm{S}_{LT}}g_T^{ij}\notag\\
&-\frac{-4\left(\bm{k}_T \cdot \bm{S}_{LT}\right)^2+2\bm{k}_T^2 \bm{S}_{LT}^2}{\bm{k}_T \cdot \bm{S}_{LT}}\frac{\bm{k}_T^i \bm{k}_T^j}{\bm{k}_T^2}\Bigg)\Phi_{LT}^{ij}\left(x,\bm{k}_T\right)\notag\\
=&-\Bigg\{-g_3^2 \bm{k}_T^2 \Big(-\bm{k}_T^2+M^2 (x-1)^2-M_S^2\Big)\Big(\bm{k}_T^2+\left(M_S+M (x-1)\right)^2\Big)\Big(\bm{k}_T^2+\left(M_S-xM+M\right)^2\Big)\notag\\
&+8 g_2^2 M^4 (x-1)\bigg[x \bm{k}_T^2 \left(M^2 (x-1)^2+2 x M_S^2\right)+(2 x-1) \bm{k}_T^4+x^2M_S^2 \left(M_S^2-M^2 (x-1)^2\right)\bigg]\notag\\
&-2 g_2 g_3 M^2\bigg[\bm{k}_T^4 \Big(-M^2 (3 x-2) (x-1)^2-\left(4 x^2+x-2\right)M_S^2\Big)+\bm{k}_T^2 \Big(M^4 (2 x-1) (x-1)^4\notag\\
&+2 M^2 (2 (x-1) x+1)(x-1)^2 M_S^2+\left(-4 x^2+2 x-1\right) M_S^4\Big)+(3-4 x)\bm{k}_T^6+x \big(M^2 (x-1)^2-M_S^2\big)^3\bigg]\notag\\
&+8 g_1 M^2 (x-1)\bigg[g_2 M^2 \Big(M_S^2 \big(\left(-4 x^2+x+1\right) \bm{k}_T^2+2M^2 x (x-1)^3\big)+M^2 (x-1)^3 \bm{k}_T^2+(3-4 x) \bm{k}_T^4\notag\\
&+M^4 x(x-1)^4+(1-2 x) x M_S^4\Big)-g_3 \bm{k}_T^2\Big(\left(\bm{k}_T^2+M_S^2\right)^2-M^4 (x-1)^4\Big)\bigg]\notag\\
&-16g_1^2 M^4 (x-1)^2 \bm{k}_T^2 \big(-\bm{k}_T^2+M^2 (x-1)^2-M_S^2\big)\Bigg\}\notag\\
&\times\Bigg\{256\pi ^3 M^4 (x-1)^2 x M_S^2 \left(\bm{k}_T^2+x \left(M^2(x-1)+M_S^2\right)\right)^2\Bigg\}^{-1}\,,\\
h_{1LT}^\perp\left(x,\bm{k}_T^2\right)=&\frac{-4M^3}{2\bm{k}_T^2 \left(\left(\bm{k}_T \cdot \bm{S}_{LT}\right)^2-\bm{k}_T^2 \bm{S}_{LT}^2 \right)}\Bigg(-S_{LT}^{\left\{i\right.} \bm{k}_T^{\left.j\right\}} + \frac{3\left(\bm{k}_T \cdot \bm{S}_{LT}\right)^2-3\bm{k}_T^2 \bm{S}_{LT}^2 }{\bm{k}_T \cdot \bm{S}_{LT}}g_T^{ij}\notag\\
&+\frac{8\left(\bm{k}_T \cdot \bm{S}_{LT}\right)^2-6\bm{k}_T^2 \bm{S}_{LT}^2}{\bm{k}_T \cdot \bm{S}_{LT}}\frac{\bm{k}_T^i \bm{k}_T^j}{\bm{k}_T^2}\Bigg)\Phi_{LT}^{ij}\left(x,\bm{k}_T\right)\notag\\
=&\Bigg\{-8 \left(2 g_1-g_2\right) M^4 (x-1)^2 \Big(g_2 \left(\bm{k}_T^2+xM_S^2\right)-g_1 \left(\bm{k}_T^2-M^2 (x-1)^2+M_S^2\right)\Big)-2 g_3M^2 (x-1)\notag\\
&\times \bigg[4 g_1 \left(\left(\bm{k}_T^2+M_S^2\right)^2-M^4(x-1)^4\right)+g_2 \Big(-2 \bm{k}_T^2 \left(M^2 (x-1)^2+(2 x+1)M_S^2\right)-3 \bm{k}_T^4+M^4 (x-1)^4\notag\\
&+2 M^2 (2 x-1) (x-1)^2 M_S^2+(1-4x) M_S^4\Big)\bigg]+g_3^2 \bigg[\bm{k}_T^4 \left(M^2 (x-1)^2+3M_S^2\right)+\bm{k}_T^2 \Big(-M^4 (x-1)^4\notag\\
&-2 M^2 (x-1)^2 M_S^2+3M_S^4\Big)+\bm{k}_T^6-\Big(M^2 (x-1)^2-M_S^2\Big)^3\bigg]\Bigg\}\notag\\
&\times\Bigg\{64\pi ^3 M^2 (x-1)^2 x M_S^2 \left(\bm{k}_T^2+x \left(M^2(x-1)+M_S^2\right)\right)^2\Bigg\}^{-1}\,.
\end{align}

The TT tensor polarized gluon TMDs are given by
\begin{align}
&f_{1TT}\left(x,\bm{k}_T^2\right)-h_{1TT}^\perp\left(x,\bm{k}_T^2\right)\notag\\
=&\frac{-M g_T^{ij}}{k_T^{\alpha \beta} S_{TT}^{\alpha \beta} }\Phi_{TT}^{ij}\left(x,\bm{k}_T\right)\notag\\
=&\Bigg\{2 g_3 M^2 \bigg[g_2 \Big(2 \bm{k}_T^2 \big(M^2 (2 x-1) (x-1)^2+(2(x-1) x+1) M_S^2\big)+(3 x-2) \bm{k}_T^4+x \left(M_S^2-M^2(x-1)^2\right)^2\Big)\notag\\
&-4 g_1 (x-1) \bm{k}_T^2 \left(\bm{k}_T^2+M^2(x-1)^2+M_S^2\right)\bigg]+g_3^2 \bm{k}_T^2 \Big(2 \bm{k}_T^2 \left(M^2(x-1)^2+M_S^2\right)+\bm{k}_T^4+\left(M_S^2-M^2(x-1)^2\right)^2\Big)\notag\\
&+4 M^4 (x-1) \bigg[g_2^2 \Big((2 x-1)\bm{k}_T^2+M^2 x (x-1)^2+x \left(2 x^2-1\right) M_S^2\Big)-2 g_1 g_2\Big((3 x-2) \bm{k}_T^2+M^2 x (x-1)^2\notag\\
&+x (2 x-1) M_S^2\Big)+4 g_1^2(x-1) \bm{k}_T^2\bigg]\Bigg\}\notag\\
&\times\Bigg\{128 \pi ^3 M^2 (x-1) x M_S^2 \left(\bm{k}_T^2+x\left(M^2 (x-1)+M_S^2\right)\right)^2\Bigg\}^{-1}\,,\\
&h_{1TT}\left(x,\bm{k}_T^2\right)\notag\\
=&\Bigg(\frac{2k_T^{ij}}{S_{TT}^{\alpha \beta}k_{T\alpha \beta}}-\frac{\bm{k}_T^4 S_{TT}^{\alpha \beta}S_{TT\alpha \beta} }{2S_{TT}^{\alpha \beta}k_{T\alpha \beta}-\bm{k}_T^4 S_{TT}^{\alpha \beta}S_{TT\alpha \beta}}\bigg(\frac{S_{TT}^{ij}}{S_{TT}^{\alpha \beta}S_{TT\alpha \beta}}+\frac{2k_T^{ij\alpha} k_{T\alpha}}{\bm{k}_T^2 S_{TT}^{\alpha \beta}k_{T\alpha \beta}} \bigg)  \Bigg) \Phi_{TT}^{ij}\left(x,\bm{k}_T\right)\notag\\
=&\Bigg\{4 g_2 M^2 \bm{k}_T^2 \bigg[g_3 \Big(\bm{k}_T^2 \big(M^2 (3 x-1)(x-1)^2+(x (2 x-1)+1) M_S^2\big)+(2 x-1) \bm{k}_T^4+x \left(M_S^2-M^2(x-1)^2\right)^2\Big)\notag\\
&-4 g_1 M^2 (x-1) \left((2 x-1)\left(\bm{k}_T^2+x M_S^2\right)+M^2 x (x-1)^2\right)\bigg]+\bm{k}_T^4\bigg[-2 g_3 M_S^2 \left(M^2 (x-1) \left(g_3 (x-1)+4g_1\right)-g_3 \bm{k}_T^2\right)\notag\\
&+\left(M^2 (x-1) \left(4 g_1-g_3(x-1)\right)-g_3 \bm{k}_T^2\right)^2+g_3^2 M_S^4\bigg]+4 g_2^2 M^4\bigg[2 x M_S^2 \left((2 (x-1) x+1) \bm{k}_T^2-M^2 (x-1)^2x\right)\notag\\
&+\left((2 x-1) \bm{k}_T^2+M^2 x (x-1)^2\right)^2+x^2M_S^4\bigg]\Bigg\}\notag\\
&\times\Bigg\{256 \pi ^3 M^4 (x-1) x M_S^2 \left(\bm{k}_T^2+x \left(M^2(x-1)+M_S^2\right)\right)^2\Bigg\}^{-1}\,,\\
&h_{1TT}^{\perp \perp}\left(x,\bm{k}_T^2\right)\notag\\
=&\frac{4M^4 S_{TT}^{\alpha \beta}S_{TT\alpha \beta} }{2S_{TT}^{\alpha \beta}k_{T\alpha \beta}-\bm{k}_T^4 S_{TT}^{\alpha \beta}S_{TT\alpha \beta}}\bigg(\frac{S_{TT}^{ij}}{S_{TT}^{\alpha \beta}S_{TT\alpha \beta}}+\frac{2k_T^{ij\alpha} k_{T\alpha}}{\bm{k}_T^2 S_{TT}^{\alpha \beta}k_{T\alpha \beta}} \bigg)   \Phi_{TT}^{ij}\left(x,\bm{k}_T\right)\notag\\
=&\Bigg\{-4 g_3 M^2 (x-1) \bigg[\left(2 g_1-g_2\right) \left(\bm{k}_T^2+M^2(x-1)^2\right)+\left(g_2 (1-2 x)+2 g_1\right) M_S^2\bigg]+g_3^2\bigg[2 \bm{k}_T^2 \left(M^2 (x-1)^2+M_S^2\right)\notag\\
&+\bm{k}_T^4+\left(M_S^2-M^2(x-1)^2\right)^2\bigg]+4 \left(g_2-2 g_1\right)^2 M^4(x-1)^2\Bigg\}\notag\\
&\times\Bigg\{64 \pi ^3 (x-1) x M_S^2 \left(\bm{k}_T^2+x \left(M^2(x-1)+M_S^2\right)\right)^2\Bigg\}^{-1}\,.
\end{align}

\section{Positivity bounds}\label{appendix2}
At leading twist, the positivity bounds are the following nine inequalities:
\begin{align}
\frac{\bm{k}_T^2}{2M^2}\left|h_1^\perp-h_{1LL}^\perp\right|&\leq f_1-f_{1LL}\,,\\
\frac{\bm{k}_T^4}{16M^4}\left[4\left(h_{1L}^\perp\right)^2+\left(2h_{1}^\perp+h_{1LL}^\perp\right)^2\right]&\leq \left(f_1+\frac{f_{1LL}}{2}+g_1\right)\left(f_1+\frac{f_{1LL}}{2}-g_1\right)\,,\\
\frac{\bm{k}_T^2}{2M^2}\left(h_1^2+4h_{1LT}^2\right)&\leq \left(f_1-f_{1LL}\right) \left(f_1+\frac{f_{1LL}}{2}+g_1\right)\,,\\
\frac{\bm{k}_T^6}{8M^6}\left[\left(h_{1T}^\perp\right)^2+\left(h_{1LT}^\perp\right)^2\right]&\leq \left(f_1-f_{1LL}\right) \left(f_1+\frac{f_{1LL}}{2}-g_1\right)\,,\\
\frac{\bm{k}_T^2}{2M^2}\left[\left(f_{1T}^\perp +g_{1LT}\right)^2+\left(f_{1LT}+g_{1T}+h_{1LT}\right)^2\right]&\leq \left(f_1-f_{1LL}\right) \left(f_1+\frac{f_{1LL}}{2}+g_1\right)\,,\\
\frac{\bm{k}_T^2}{2M^2}\left[\left(f_{1T}^\perp -g_{1LT}\right)^2+\left(f_{1LT}-g_{1T}+h_{1LT}\right)^2\right]&\leq \left(f_1-f_{1LL}\right) \left(f_1+\frac{f_{1LL}}{2}-g_1\right)\,,\\
\left|h_{1TT}\right|&\leq \frac{1}{2}\left(f_1+\frac{f_{1LL}}{2}+g_1\right)\,,\\
\frac{\bm{k}_T^4}{2M^4}\left|h_{1TT}^{\perp \perp}\right|&\leq \left(f_1+\frac{f_{1LL}}{2}-g_1\right)\,,\\
\frac{\bm{k}_T^4}{M^4}\left[g_{1TT}^2+\left(f_{1TT}-h_{1TT}^\perp\right)^2\right]&\leq \left(f_1+\frac{f_{1LL}}{2}+g_1\right)\left(f_1+\frac{f_{1LL}}{2}-g_1\right)\,.
\end{align}

For integrated TMDs case, there are the following three bounds:
\begin{align}
\left|g_1\right|&\leq f_1 +\frac{f_{1LL}}{2}\,,\\
f_{1LL}&\leq f_1\,,\\
\left|h_{1TT}\right|&\leq \frac{1}{2} \left(f_1+\frac{f_{1LL}}{2}+g_1\right)\,.
\end{align}

\end{widetext}

\end{document}